\documentclass[aps,prd,twocolumn,superscriptaddress,groupedaddress,nofootinbib]{revtex4}
\usepackage{graphicx}  
\usepackage{dcolumn}   
\usepackage{bm}        
\usepackage{amssymb}
\usepackage{amstext}
\usepackage{amsmath}
\usepackage{dsfont}
\usepackage{xcolor}
\usepackage{amsthm}
\usepackage{enumitem}
\usepackage{tensor}
\usepackage[colorlinks]{hyperref}
\hypersetup{
     breaklinks=true,
    pdfstartview={FitH},    
    colorlinks=true,       
    linkcolor=blue,          
    citecolor=red,        
    filecolor=magenta,      
    urlcolor=blue,           
    anchorcolor=green,      
    linktocpage=true
}

\newcommand{\be}{\begin{equation}}\newcommand{\ee}{\end{equation}}
\newcommand{\bea}{\begin{eqnarray}}\newcommand{\eea}{\end{eqnarray}}
\newcommand{\brr}{\begin{array}}\newcommand{\err}{\end{array}}
\newcommand{\bit}{\begin{itemize}}\newcommand{\eit}{\end{itemize}}
\newcommand{\ben}{\begin{enumerate}}\newcommand{\een}{\end{enumerate}}

\newcommand{\ba}{\begin{array}}
\newcommand{\ea}{\end{array}}

\definecolor{darkred}{rgb}{.8,0,0}

\definecolor{darkblue}{rgb}{0,0,0}
\newcommand{\cdblue}{\color{darkblue}}

\def\1{{_{1}}}\def\2{{_{2}}}

\def\noHe0{:\;\!\!\;\!\!:H_e(0):\;\!\!\;\!\!:}
\def\noHm0{:\;\!\!\;\!\!:H_\mu(0):\;\!\!\;\!\!:}
\begin{document}
\title{Generalized Birkhoff theorems and 2+2 direct pruduct spacetimes in Weyl conformal gravity}
\author{Petr Jizba}
\email{p.jizba@fjfi.cvut.cz}
\affiliation{FNSPE,
Czech Technical University in Prague, B\v{r}ehov\'{a} 7, 115 19, Prague, Czech Republic}
\author{Tereza Lehe\v{c}kov\'{a}}
\email{lehecter@fjfi.cvut.cz}
\affiliation{FNSPE,
Czech Technical University in Prague, B\v{r}ehov\'{a} 7, 115 19, Prague, Czech Republic}
\date{\today}
\begin{abstract}
In this paper, we study 2+2 direct product spacetimes sourced by separated electromagnetic and Yang--Mills fields within Weyl conformal gravity. We prove that all such configurations admit at least 2 independent, commuting non-null Killing vectors, which we use to find general solutions. As a special case, we obtain a generalization of the Birkhoff--Riegert theorem to all spacetimes containing a two-dimensional subspace of constant Gaussian curvature, and we also revisit the original formulation of the theorem.  We further analyze the resulting solutions in terms of Weyl equivalence classes. Their connections to known solutions in both Weyl conformal gravity and Einstein gravity are established through conformal relations. We also examine the fundamental physical and geometric properties of the newly obtained configurations and their equivalence classes.

\end{abstract}

\vskip -1.0 truecm

\maketitle

\section{Introduction}
\label{uvod}

Birkhoff's theorem (BT), a cornerstone result of general relativity (GR), states that any spherically symmetric solution of the vacuum Einstein equations is uniquely described by the Schwarzschild metric
\begin{equation}
ds^{2}=-\left(1-\frac{2m}{r}\right)dt^{2}+\left(1-\frac{2m}{r}\right)^{-1}dr^{2}+r^{2}d\Sigma^{2}\, ,  
\label{I.1.kl}
\end{equation}
where $d\Sigma^2$ is a shorthand for $\left(d\theta^2 \ + \ \sin^2\theta d\varphi^2\right)$.  
Equivalently, BT implies that every spherically symmetric vacuum spacetime (ST) in GR necessarily admits an additional symmetry. In Schwarzschild coordinates, this is manifested by the existence of the Killing vector field (KVF) $\partial_{t}$. To establish a correct Newtonian limit, $m$ in~(\ref{I.1.kl}) is interpreted as a (positive) mass of the source; therefore, the horizon arises and its outer region is static~\cite{Birkhoff,Eiesland,Jebsen,Schleich,Ehlers}. Furthermore, the inclusion of an electromagnetic (EM) field or a cosmological constant preserves the presence of the additional symmetry. This property is closely related to black hole uniqueness theorems~\cite{BHun}.
Since BT implies that the external gravitational field of a spherically symmetric mass distribution is necessarily time independent --- regardless of the internal dynamics of the source, provided the region under consideration is vacuum --- it follows that spherically symmetric oscillations cannot generate gravitational waves. This fact imposes a significant restriction in the framework of GR.

BT has far-reaching implications and is commonly employed in both astrophysical and cosmological contexts~\cite{Straumann,Padmanabhan}. 
For instance, it guarantees the stability and time-independence of the ST outside a non-rotating star or black hole~\cite{Wald:84}. It also ensures that the gravitational field inside a spherically symmetric shell of matter is flat, in direct analogy with Newton’s shell theorem~\cite{MTW}, etc.

Some modified gravity theories (MGTs) admit Birkhoff-like theorems, but typically only when additional constraints are imposed. In other words, the uniqueness and static character of spherically symmetric solutions in the modified vacuum (i.e., a vacuum defined by the modified field equations) can be established, although often only under specific assumptions. Examples include $f(R)$ gravity models~\cite{Sotiriou:2007zu,Capozziello:08,Faraoni:2010rt} or scalar–tensor theories with minimally coupled scalar fields where the scalar field is constant in the vacuum region~\cite{Krori:77}. 
{\cdblue{In other MGTs, such as Brans--Dicke theory, BT fails~\cite{Capozziello:12}, whereas in higher-order curvature theories, including Gauss--Bonnet and Lovelock gravity in higher dimensions, the situation is more subtle~\cite{Bogdanos,Camanho:2015}. For instance, spherically symmetric vacuum solutions in Lovelock gravity are determined by an algebraic equation that typically admits multiple branches, each corresponding to a distinct ST~\cite{Boulware:1985,Wheeler:1986a,Wheeler:1986b}. This multi-branch structure allows for nontrivial dynamical phenomena, such as spherically symmetric shock waves~\cite{Gravanis:10}, and implies that uniqueness and staticity hold only within each branch. Nevertheless, suitably generalized versions of BT can be formulated for Lovelock gravity under appropriate conditions~\cite{Camanho:2015}.}} In general, the loss or {\cdblue{generalization}} of BT signals that the theory admits richer spherically symmetric dynamics, 
which may offer both observational opportunities and theoretical challenges.

An important class of MGTs involves higher-derivative extensions, with Weyl conformal gravity (WCG) as a notable axample~\cite{Weyl}. WCG exhibits rich classical phenomenology~\cite{MK a dalsi,making case,mannheim_b,mannheim_c,MK2,Kazanas,mannheim,mannheim_a,JKS2,Edery:1998zi,Edery:1997hu,Edery:2001at,PJ-GL-23,NP} and  has been extensively studied for its ability to address the dark matter problem through its observational predictions about the large-scale dynamics of the Universe (galactic and beyond)~\cite{Diaferio,Varieschi}. Additionally, unlike GR, WCG is perturbatively renormalizable when quantized~\cite{Stelle:77}, with a host of new phenomena, including  gravitational instantons (that encode information
about a non-perturbative vacuum structure of  quantum WCG), an antiscreening behavior at high energies due to the negative $\beta$-function, (perturbative) ghosts, etc.~\cite{fradkin:85,Strominger:85,Hartnoll:06,Taubes:88,kWCG1}. Within the framework of classical WCG, BT  plays an interesting role with a number of intriguing consequences.

In the context of WCG, a key result is the analogue of BT established by Riegert~\cite{Riegert}, which we shall refer to as the Birkhoff--Riegert theorem (BRT) in the following.
In his proof, Riegert considered a general spherically symmetric metric in a  2$+$2 decomposition gauge, with an EM field sharing the same structure and symmetry, and showed that such configurations always admit a non-trivial, non-null KVF which can then be used to adjust the metric to a suitable form and find general solution. This, in turn, implied that any other spherically symmetric solution can be derived from the 2$+$2 case and, furthermore, is equivalent to the well-known WCG solution of Mannheim and Kazanas (MK) or to its electrically charged generalization~\cite{MK a dalsi}.

Although the main idea behind Riegert's proof is formally sound, the argument appears to rely on implicit assumptions that warrant further clarification and seems to contain several minor inaccuracies.
We seek to clarify the ensuing in some detail in the following sections. Most importantly, Riegert's argument does not account for situations when the Weyl conformal factor degenerates, i.e. becomes singular or vanishes.  This is reminiscent to a situation known from scalar-tensor gravity theories, such as Brans--Dicke theory or Higgs inflation, which can be described in either the Jordan or the Einstein frames~\cite{capozziello_faraoni2010}. The two frames are related by local Weyl (or conformal) rescaling of the metric: $g_{\mu\nu}^{(E)} = \Omega^2(\phi)\ \! g_{\mu\nu}^{(J)}$. While such a Weyl transformation preserves the light-cone structure, and thus the \emph{local} causal structure (i.e., which events can influence which others, and the distinction between lightlike and timelike separation) is identical in both frames, the \emph{global} properties of the two STs can be fundamentally different if the Weyl factor $\Omega(x)$ is not regular everywhere. For example, we will see that singularities, conformally flat parts of ST, or horizons can emerge or vanish when degenerate Weyl transformation is applied. So, generally, not all spherically symmetric solutions in WCG are globally equivalent to MK solutions --- they might be both causally and topologically distinct. As with any gauge theory, WCG should be naturally formulated in terms of distinct gauge equivalence classes (characterized by their symmetry, e.g. spherical). However, only solutions that share the same causal and topological structure can be regarded as belonging to a single gauge equivalence subclass, within which it is possible to shift via non-degenerate Weyl transformation.

For more general STs, the 2$+$2 direct product decomposition approach with some modifications was used by Dzhunushaliev and Schmidt~\cite{HJS1} (see also references therein). They demonstrated that vacuum 2$+$2 STs in WCG always admit two independent, non-null KVFs (a result they referred to as the ``double Birkhoff theorem'') and, building on Schmidt’s earlier analysis of two-dimensional gravity~\cite{HJS2D}, used these symmetries to obtain general solutions. Consequently, they also derived the most general form of WCG STs conformal to 2$+$2 direct product ones. Nevertheless, these analysis omitted the inclusion of EM (or other) fields, did not consider the connections between STs related by Weyl transformations, and left the physical properties of the solutions largely unexplored. It should be emphasized, however, that numerous other works on WCG STs belonging to this class exist, focusing mostly (but not exclusively) on spherically symmetric solutions and their potential physical implications~\cite{MK a dalsi, bubles, NP, EF, TBH, C_WCG, NWmass2, tdcervi}.

This paper aims to extend the results of Riegert and Schmidt. Specifically, we refine their analyzes by properly accounting for degenerate conformal factors and examining the corresponding equivalence classes of STs. Furthermore, we extend the main results of Ref.~\cite{HJS1}  by including contributions from both EM and Yang--Mills (YM) fields. Finally, we analyze the fundamental properties of the obtained solutions and examine their connections with other well-known solutions in both WCG and GR.

The structure of the paper is as follows. In Sec.~\ref{basics}, we outline the general framework, providing the basic definitions, relations, solutions, and transformation rules used in WCG. We also include a brief discussion of maximally symmetric 2D spaces. In Sec.~\ref{theorem}, we generalize Schmidt's and Riegert's procedure for constructing KVFs with EM and YM fields present. We then employ these KVFs to transform the metrics into a canonical form and derive the corresponding solutions. 
We also examine the connection to BRT and provide a revised analysis of its formulation.
Finally, we propose a refined classification scheme based on causally equivalence classes and subclasses.
In Sec.~\ref{properties}, we examine the basic properties of the obtained solutions, including geometrical characteristics such as Weyl and Ricci scalars, alternative coordinate systems, and horizons. We also discuss their relation to the corresponding solutions in GR --- highlighting explicit conformal relations --- as well as their connection to previously known solutions in WCG. Finally, in Sec.~\ref{diskuse}, we present a discussion of the results obtained and outline possible directions for future research. 
For the reader's convenience, the paper is supplemented with two appendices. Appendix~\ref{apA} provides a list of acronyms used throughout the main text, while Appendix~\ref{apB} contains additional technical material related to conformal Einstein spaces.

\section{Basic definitions and relations}
\label{basics}

\subsection{WCG and Weyl transformations}
\label{WCG}

In this section, we provide only the most essential elements of the theory; for a comprehensive treatment, we refer the reader to~\cite{kWCG1, kWCG2, making case}. 
WCG is a purely metric theory of gravity that possesses the standard diffeomorphism invariance of GR, supplemented by the invariance under Weyl (or conformal) transformations
\begin{equation}
    g_{\mu\nu} \ \rightarrow\  \tilde{g}_{\mu\nu} \ = \ \Omega^2(x)\ \! g_{\mu\nu}  \, .
    \label{II.2.cg}
\end{equation}
In four dimensions, the form of the action is
\begin{eqnarray}
 S_{_{\rm W}}[g] &=&-\frac{1}{4G_{_{\!\rm W}}^2} \int dx^4 \sqrt{\vert g \vert} \ \! C^{\mu \nu \lambda \kappa} C_{\mu \nu \lambda \kappa}\nonumber \\[2mm]
  &=&  -\frac{1}{2G_{_{\!\rm W}}^2} \int dx^4 \sqrt{\vert g \vert} \ \! \Big( R_{\mu \nu}R^{\mu \nu} \ - \ \tfrac{1}{3}R^2\Big)\, ,~~~~
  \label{akce}
\end{eqnarray}
where $G_{_{\rm W}}$ is a constant and $C_{\mu \nu \lambda \kappa}$ is a Weyl tensor
\begin{eqnarray}
C_{\mu\nu\rho\sigma}  &=&
R_{\mu\nu\rho\sigma} \ - \ \left(g_{\mu[\rho}R_{\sigma]\nu} -
g_{\nu[\rho}R_{\sigma]\mu} \right)
\nonumber\\[1mm]
&+& \frac{1}{3}R \ \! g_{\mu[\rho}g_{\sigma]\nu}\, .
\label{Weylcomp}
\end{eqnarray}
Here, brackets around the indices refer to the antisymmetric part.
The first and second forms in~(\ref{akce}) are equivalent up to the topological (boundary) term. The field equations resulting from this action are called Bach vacuum equations and are of the form
\begin{equation}
\label{Bach}
    B_{\mu \nu} \ \equiv \ \left(\nabla^\kappa \nabla^\lambda +\tfrac{1}{2}R^{\kappa \lambda}\right)C_{\mu \kappa \nu \lambda} \ = \ 0\, ,
\end{equation}
where $B_{\mu \nu}$ is the so-called Bach tensor, which is symmetric and trace-free, and transforms under the Weyl transformation as $B^{\mu \nu} \rightarrow B^{\mu \nu} \Omega^{-6}$. In the presence of matter (i.e. stress-energy tensor), the equations read 
\begin{equation}
\label{Bachnonvac}
B_{\mu \nu} \ = \  \frac{G_{_{\!\rm W}}^2}{2} T_{\mu \nu}\, ,
\end{equation}
which places some non-trivial restrictions on the allowed energy-momentum tensor forms --- they must be traceless, and so the field-matter part of the action must be Weyl invariant. 

In this paper, we work with the EM stress–energy tensor
\begin{equation}
\label{EMt}
T^{\rm{EM}}_{\mu \nu} \ = \ \frac{1}{e^2}\Big(F_{\mu \lambda}F_\nu^\lambda-\tfrac{1}{4}g_{\mu \nu}F_{\alpha \beta}F^{\alpha \beta}\Big)\, ,
\end{equation}
where $e$ is the gauge coupling (the electric charge in electrodynamics). This form of the stress–energy tensor follows from the standard (Weyl invariant) EM action
\begin{equation}
\label{EMa}
S_{\rm{EM}} \ = \ -\frac{1}{4e^2}\int d^4x\sqrt{\vert g \vert} \ \! F_{\mu \nu}F^{\mu \nu}\, , 
\end{equation}
and thus automatically meets the key requirements for a valid stress–energy tensor in WCG, namely, it is symmetric and trace-free. Ensuing generalization to YM fields will be discussed in Sec.~\ref{YM}.

The fourth-order structure of the Bach field equations~(\ref{Bach}) complicates their mathematical analysis, though various powerful techniques such as the Newman--Penrose or Geroch--Held--Penrose formalisms can greatly facilitate this task~\cite{NP,Contreras,Durkee}.  Importantly, WCG offers new insights into some of the central challenges of GR, such as the cosmological constant, the dark sector, singularity issues, and renormalizability, thereby representing a potentially viable gravitational framework (see, e.g., Refs.~\cite{MK a dalsi,making case,mannheim_b,mannheim_c,MK2,Kazanas,mannheim,mannheim_a,JKS2,Edery:1998zi,Edery:1997hu,Edery:2001at,PJ-GL-23,NP,Diaferio,Varieschi,Stelle:77}).

By employing the contracted Bianchi identity,
\begin{equation}
\nabla^{\delta} C_{\alpha \beta \gamma \delta} \ = \  - \nabla_{[\alpha} \left(R_{\beta]\gamma} \ - \  \tfrac{1}{6} R g_{\beta]\gamma}\right)\, ,
\end{equation}
the Bach tensor can be rewritten in terms of the Schouten tensor (see. e.g.,~\cite{Besse})
\begin{eqnarray}
P_{\alpha\beta} \ = \  \frac{1}{2} \left(R_{\alpha\beta} \ - \  \tfrac{1}{6} R g_{\alpha\beta}\right)\, ,
\end{eqnarray}
as  
\begin{eqnarray}
&&B_{\mu\nu} \ = \  2 \nabla^{\kappa} \nabla_{[\kappa} P_{\mu]\nu} \ + \  P^{\kappa \lambda} C_{\mu\kappa\nu\lambda}\, . 
\end{eqnarray}
In this connection, we might note that for pure gravity with cosmological constant $\Lambda$, Einstein's field equations read
\begin{equation}
R_{\alpha \beta} \ = \  {\Lambda} g_{\alpha \beta} \;\;\;\; \Rightarrow \;\;\;\;  R \ = 
4 {\Lambda}\, .
\label{II.10.cg}
\end{equation}
Solutions of~(\ref{II.10.cg}) define the so-called Einstein spaces. In such spaces, \(P_{\alpha\beta} =  g_{\alpha\beta} \ \! {\Lambda}/6\), which directly implies that all Einstein spaces automatically also satisfy the vacuum Bach equation. The reverse implication, however, is not true.
In fact, there even exist vacuum solutions of the vacuum Bach equation that are not Weyl-equivalent to any ST of GR~\cite{HJS2}. WCG thus permits genuinely new, non-Einstein solutions. Even in cases when WCG admits analogues of familiar GR STs, important differences can arise~\cite{MK2,Kazanas,mannheim,mannheim_a}.
A celebrated example of a non-Einstein solution is the MK solution~\cite{MK a dalsi}, i.e., the WCG analogue of the Schwarzschild--(anti-)de Sitter solution, which is given by the line element
\begin{eqnarray}
ds^2 \ = \  -H(r)dt^2 \ + \ H(r)^{-1}dr^2  \ + \ r^2 d\Sigma^2\, , 
    \label{MK}
\end{eqnarray}
where 
\begin{eqnarray}
&&\mbox{\hspace{-5mm}}H(r) \ = \ 1 \ - \ 3\beta\gamma \ - \ \frac{\beta(2  -  {\cdblue{3}}\beta \gamma)}{r} \  + \ \gamma r\ - \ \kappa r^2\, .~~~~
\end{eqnarray}
Here we note that $\gamma$, which is not present in GR can be interpreted as a dark-matter factor, $\kappa$ as a cosmological constant and $\beta$ as the mass of the source \cite{MK a dalsi}. In the charged case (i.e., when a Coulombic field is included), the function 
$H(r)$ acquires an additional contribution of order $1/r$, 
in contrast to the GR counterpart, where the corresponding  term is of order $1/r^{2}$. It is also possible, though cosmologically less feasible, to add magnetic charge in the same manner~\cite{MK a dalsi}. In the vacuum case, the linear term can be transformed out, changing the metric to a Schwarzchild--(anti-)de Sitter metric~\cite{NWmass2, HJS3}. However, this transformation has singular points, so the STs obtained cannot be considered globally equivalent, as we will discuss later. 

To proceed, we introduce the transformation rules needed for later sections.  In particular, since we will use the tetrad formalism~\cite{NP,kniha}, we start with the associated transformations. 
In this framework, four null vectors, ${l, n, m, \bar{m}}$, are introduced such that they are mutually orthogonal, except for the inner products $l_\mu n^\mu = -1$ and $m_\mu \bar{m}^\mu = 1$. The metric can then be expressed as
\begin{equation}
g_{\mu \nu} \ = \  -l_\mu n_\nu \ - \  n_\mu l_\nu \ + \  m_\mu \bar{m}_\nu \ + \  \bar{m}_\mu m_\nu\, .
\end{equation}
Under a Weyl rescaling, the tetrad transforms as  
\begin{equation}
l^\mu \rightarrow \Omega^{-1} l^\mu\, , \qquad l_\mu \rightarrow \Omega\, l_\mu\, ,
\end{equation}
with analogous transformations for the other tetrad vectors. From the tetrads, the Weyl scalars are obtained, which transform as
\begin{equation}
\Psi_i \ \rightarrow \ \Omega^{-2} \Psi_i ,
\end{equation}
while the Ricci scalars $\Phi_{ij}$ transform in a more involved way. Consequently, the Petrov algebraic type is preserved under Weyl transformations. From Weyl scalars, several scalar polynomial invariants can be construed that reveal  scalar polynomial curvature singularities in solutions. Here we will use, for instance, invariant
\begin{equation}
\label{ID}
I \ = \ \Psi_0 \Psi_4 \ - \ 4\Psi_1 \Psi_3 \ + \ 3 \Psi_2^2\, ,
\end{equation}
while another commonly used invariant is the square of the Weyl tensor, $C^2$. As $I$ depends quadratically on the Weyl scalars $\Psi_i$, it transforms as $I \rightarrow \Omega^{-4} I$. This implies that, under a generic Weyl transformation, the magnitude of $I$ changes, but more importantly, divergences (corresponding to scalar curvature singularities) or zeros (indicating locally conformally flat regions of the ST) may be introduced or eliminated by a degenerate transformation. Here, {\em degenerate} refers to a Weyl transformation in which the conformal factor $\Omega$ either vanishes or diverges at some points.

Another set of objects of central importance in our analysis are KVFs and conformal Killing vector fields (CKVFs). To study their behavior under Weyl transformations, we first recall how the Christoffel symbols transform under~(\ref{II.2.cg}). Explicitly, we have  
\begin{widetext}
\begin{equation}
\tilde{\Gamma}^\iota_{\kappa\lambda} 
\ = \  \tilde{g}^{\iota\mu} \big( \tilde{g}_{\mu \kappa , \lambda} \ + \ \tilde{g}_{\mu \lambda , \kappa} \ - \  \tilde{g}_{\kappa \lambda , \mu} \big) 
\ = \ \Omega^{-2} g^{\iota\mu} \big[ (\Omega^2 g_{\mu \kappa})_{,\lambda} \ + \  (\Omega^2 g_{\mu \lambda})_{,\kappa} \ - \  (\Omega^2 g_{\kappa \lambda})_{,\mu} \big]\, .
\end{equation}
Simplifying, this yields
\begin{equation}
\tilde{\Gamma}^\iota_{\kappa\lambda} 
\ = \ \Gamma^\iota_{\kappa\lambda} \ + \ \Omega^{-1} \big( \delta^\iota_\kappa \, \Omega_{,\lambda} \ + \ \delta^\iota_\lambda \, \Omega_{,\kappa} \ - \  g^{\iota\mu} g_{\kappa \lambda} \, \Omega_{,\mu} \big)\, , 
\end{equation}
where $\delta^\iota_\kappa = g^{\iota\mu}  g_{\mu \kappa}$ is the Kronecker delta. 
With these results, we can explicitly transform the (C)KVF equation and verify that a CKVF remains a CKVF under Weyl transformations (even if the transformation is degenerate). In contrast, a KVF may turn into a CKVF under Weyl transformations. Consider $V$ as a CKVF for the metric $g_{\mu \nu}$. In order to show that $V$ continues to be a CKVF for the Weyl rescaled metric $\tilde{g}_{\mu \nu}$, we define $\tilde{V}^\mu = V^\mu$ (which directly implies that $\tilde{V}_\mu = \Omega^2 V_\mu$) and write
\begin{eqnarray}
\tilde{V}_{\mu;\nu} \ + \ \tilde{V}_{\nu;\mu} \ &=& \ \tilde{V}_{\mu,\nu} \ + \ \tilde{V}_{\nu,\mu} \ - \ 2\tilde{\Gamma}^{\alpha}_{\mu \nu}\tilde{V}_{\alpha} \nonumber \\[2mm] 
&=& (\Omega^{2}V_{\mu})_{,\nu} \ + \ (\Omega^{2}V_{\nu})_{,\mu}\ - \ 2\left[\Gamma^{\alpha}_{\mu \nu}+\Omega^{-1}(\delta^\alpha_{ \mu}\Omega_{,\nu}\ + \ \delta^\alpha_{ \nu}\Omega_{,\mu} \ - \ g^{\alpha \gamma}g_{\mu \nu}\Omega_{,\gamma})\right]\Omega^{2}V_{\alpha}
\nonumber \\[2mm] &=&
\Omega^{2}\left(V_{\mu, \nu} \ + \ V_{\nu, \mu} \ - \ 2\Gamma^{\alpha}_{\mu \nu}V_{\alpha}\right) \ + \ 2 \Omega(V_{\mu} \Omega_{,\nu} \ + \ V_{\nu} \Omega_{,\mu})\ - \ 2 \Omega\left(V_{\mu} \Omega_{,\nu}\ + \ V_{\nu} \Omega_{,\mu} \ - \ g_{\mu \nu} V^\gamma \Omega_{,\gamma}\right) 
\nonumber \\[2mm]&=& \Omega^{2}\left(V_{\mu ;\nu} \ + \ V_{\nu ;\mu}\right) \ + \   2\Omega \ \! V^\gamma \Omega_{,\gamma} \ \! g_{\mu \nu}\, .
\label{CKVF eq. tr.}
\end{eqnarray}
\end{widetext}
When $V$ is a CKVF, the first term of the final expression can be written as $g_{\mu \nu} f(x^\mu)$, where $f$ is a scalar function of the coordinates. The second term already appears in this form. Thus, indeed CKVF$~\rightarrow~$CKVF. If, on the other hand, $V$ is a KVF, the first term vanishes. In this case, thus, KVF$~\rightarrow~$CKVF. These transition rules for (C)KVFs naturally extend  to hidden symmetries such as (conformal) Killing tensors (CKT) of arbitrary order~\cite{CKT}.

(C)KVFs are significant not only because they serve as generators of (conformal) isometries but also due to their role in the study of horizons, which are key features of STs. A (conformal) Killing horizon is defined by the condition 
\begin{eqnarray}
V_\mu V^\mu \ = \  0\, ,
\end{eqnarray}
for a (C)KVF $V$. Under a conformal transformation, this condition becomes
\begin{eqnarray}
\tilde{V}_\mu \tilde{V}^\mu  \ = \  \Omega^2 V_\mu V^\mu  \ = \  0\, .
\end{eqnarray}
If the Weyl factor $\Omega$ is regular, the location of the horizons remains unchanged. However, if $\Omega$ is degenerate, new roots may appear, or existing ones can be eliminated.

\subsection{2D constant Gaussian curvature spaces}
\label{const_curv}

Now, let us shift our focus slightly. Since maximally symmetric two-dimensional spaces (2-spaces) of constant Gaussian curvature will be important for our analysis, we summarize briefly their forms, basic features, and notation. For further details, see, e.g., Ref.~\cite{kniha}.

We begin with Euclidean 2-spaces. There are three well-known cases corresponding to constant Gaussian curvature: the sphere, the plane, and the hyperboloid, associated with positive, zero, and negative curvature, respectively. In a unified representation, they can be written as 
\begin{equation}
   d\Sigma_E^{2} \ = \ \frac{d \eta^{2}+d\lambda^{2}}{\left(1+\frac{K(\eta^{2}+\lambda^{2})}{4}\right)^{2}} \ = \ \frac{2d\zeta d\bar{\zeta}}{\left(1+\frac{K\zeta \bar{\zeta}}{2}\right)^{2}}\, , 
   \label{2DE}
\end{equation}
where $K$ is the Gaussian curvature, $K=\epsilon/a^{2}$ and $ a$ is a real constant. Here, $\epsilon$ is $0, -1$ or $1$ and $ \zeta=(\eta+i\lambda)/\sqrt{2a^{2}} $.  The KVFs corresponding to the second form of the metric~(\ref{2DE}) are 
\begin{eqnarray}
\label{X spatial}
X_E \ &=& \ (c_1K\zeta^{2}\ + \ c_2 \zeta \ + \ 2c_3)\partial_{\zeta}\nonumber \\[2mm]
&&+ \ (2c_1 \ - \ c_2\bar{\zeta} \ + \ c_3K\bar{\zeta}^{2})\partial_{\bar{\zeta}}\, .
\end{eqnarray}
Here, the $c_i$ are arbitrary constants, implying that there are three independent KVFs. 

For Lorentzian 2-space, there are again three standard cases of constant Gaussian curvature: Minkowski (zero curvature), de Sitter (positive curvature), and anti-de Sitter (negative curvature). These geometries can be represented in a compact form, for example, as
\begin{equation}
\label{2DL}
d\Sigma_L^{2} \ = \ -\frac{2du dv}{\left(1-\frac{Ku v}{2}\right)^{2}}\, ,
\end{equation}
where $K$ is defined in the same way as in the Euclidean case, with KVFs
\begin{eqnarray}
\label{X lorentzian}
X_L \ &=& \-(c_1 K u^{2} \ + \ c_2u \ + \ 2c_3)\partial_u\nonumber \\[2mm] 
&&+ \ (2c_1 \ + \ c_2v \ + \ c_3Kv^{2})\partial_v\, ,
\end{eqnarray}
where again $c_{i}$ are arbitrary.

In the four-dimensional case, nine distinct Lorentzian combinations of the above metrics can be constructed. As we will encounter them later, we briefly summarize their main features here.

Within GR, there exist three vacuum solutions, all specified by the condition $K_E = K_L = \Lambda$, where the sub-indices $E$ and $L$ denote the Euclidean and Lorentzian parts, respectively. These vacuum solutions correspond to Minkowski, Nariai, and anti-Nariai STs. The remaining combinations correspond to non-vacuum solutions, and of these, only three --- ($\epsilon_L = -1, \epsilon_E = 1$), ($\epsilon_L = 0, \epsilon_E = 1$), and ($\epsilon_L = -1, \epsilon_E = 0$) --- are considered to be physically relevant in GR, as they lead to positive values of $\Phi_{11}$ (required by the weak energy condition). These include the Bertotti--Robinson and Plebański--Hacyan STs~\cite{kniha}.

The vacuum structure of these combinations in WCG was examined in Ref.~\cite{HJS1}. Inserting the explicit metric forms into the Bach equations~(\ref{Bach}) yields five possible vacuum solutions, all satisfying $K_L^2 = K_E^2$. This condition encompasses two cases that correspond to non-vacuum geometries in GR. Furthermore, whenever a constant Gaussian curvature 2-space is embedded within the four-dimensional ST, the 
metric can be Weyl rescaled so that the curvature parameter $K$ can be directly identified with  $\epsilon$, except in cases where two constant-curvature sectors appear 
simultaneously. We will adopt this convention throughout.

\section{Generalizing Schmidt and Rigert}
\label{theorem}

In this section, we generalize the analysis of Dzhunushaliev and Schmidt~\cite{HJS1}, who demonstrated that 2$+$2 direct product STs necessarily admit a two-dimensional isometry group generated by two KVFs, and derived the corresponding general solutions.
We further extend  this analysis by incorporating EM and YM fields. Along the way, we also address other salient issues, including Riegert's treatment of BT~\cite{Riegert} and the existence of globally non-equivalent classes of STs.

\subsection{2+2 metrics and KVFs}
\label{hlavni cast}

We will work with the general 2$+$2 direct product metric
\begin{eqnarray}
ds^{2}&=&g_{ab}(x^{0}, x^{1})dx^{a}dx^{b} \ + \ g_{cd}(x^{2}, x^{3})dx^{c}dx^{d}  \nonumber \\[2mm]
&\equiv& ds_{(1)}^{2}\ + \ 
 ds_{(2)}^{2}\, .
\label{2+2m}
\end{eqnarray}
We stress that the metric form (\ref{2+2m}) should not be interpreted as a mere ansatz. Instead, it embodies a hidden symmetry requirement: because every metric is trivially a KT of itself, the decomposition in (\ref{2+2m}) implies that both $g_{ab}$ and $g_{cd}$ act as KTs independently. Therefore, we are not dealing with an arbitrary parameterization but with a well-defined symmetry class of STs.

Our goal is to demonstrate that in WCG such a metric necessarily admits two independent and commuting KVFs. For consistency, we also assume that any accompanying EM field tensor takes a separable form, $\mathcal{F}_{(1)} + \mathcal{F}_{(2)}$, i.e., in a differential 2-form language we can write
\begin{eqnarray}
\mathcal{F} \ &=& \ \frac{1}{2}F_{a b}(x^0, x^1)dx^a \wedge dx^b \nonumber \\[2mm]
 &+& \ \frac{1}{2}F_{cd}(x^2, x^3)dx^c \wedge dx^d\, .  
 \label{polesep}
\end{eqnarray}
This symmetry-adapted form is a clear analogue of an Einstein--Maxwell product geometry used in GR. For instance, STs like Bertotti--Robinson ($AdS_2 \times S^2$) or Nariai ($dS_2 \times S^2$) use exactly this kind of separation~\cite{kniha,Bertotti,Robinson}.

In Eqs.~(\ref{2+2m})-(\ref{polesep}) we have introduced the convention that $a,b \in \{0,1\}$ and $c,d \in \{2,3\}$, while Greek indices continue to range from $0$ to $3$. In what follows, whenever different Latin ST indices are employed, they will refer to the specific two-dimensional sector under discussion. Because the metric contains no mixed terms, the Riemann and Ricci tensor, Ricci scalar, and metric determinant factorize into separated contributions, i.e.
\begin{eqnarray}
&&R_{\mu\nu \kappa \lambda} \ = \  R_{\mu\nu \kappa \lambda (1)} \ + \  R_{\mu\nu \kappa \lambda(2)}\, ,\nonumber \\[2mm] 
&&R_{\mu\nu} \ = \  R_{\mu\nu (1)} \ + \  R_{\mu\nu (2)}\, , 
\;\;\;\;
R \ = \  R_{(1)} \ + \  R_{(2)}\, , \nonumber \\[2mm]
&&
\sqrt{|g|} \ = \  \sqrt{|g_{(1)}|}\,\sqrt{|g_{(2)}|}\, .
\label{III.30.aa}
\end{eqnarray}
The components of the Ricci and Riemann tensors are nonzero only if all indices of a given component belong to the same 2D sector of the metric. A similar separation property as~(\ref{III.30.aa}) holds for many other geometric quantities, such as the Christoffel symbols.

We now employ this framework to evaluate Bach tensor~(\ref{Bach}). To do so, we will also need the following 2D identities \cite{kniha2}:
\begin{equation}
\label{Ricci2D}
R_{ij} \ = \  \tfrac{1}{2}R\, g_{ij}\, ,
\end{equation}  
(i.e. all 2D manifolds are Einstein spaces)  and
\begin{equation}
\label{Riemann2D}
R_{ijkl} \ = \  \tfrac{1}{2}R\big(g_{ik}g_{jl}\ - \ g_{il}g_{jk}\big)\ .
\end{equation} 
In the 2$+$2 direct product STs, the structure of the metric allows for a considerable simplification of the Weyl tensor~(\ref{Weylcomp}). In fact, from its definition it is immediately evident that, in the 2$+$2 direct product case, any component $C_{\mu \nu \kappa \lambda}$ with three indices belonging to one 2D sector of the metric and the fourth index to the other sector vanishes. Similarly, all components of the Bach tensor~(\ref{Bach}) that mix indices from both 2D sectors are identically zero. Consequently, we are left with the components $B_{ij}$ for which both indices $i,j$ belong to the same 2D subspace. To be specific, let us assume they belong to the first subspace, which we label by $A$ and $B$. As discussed above, these components will then simplify to
\begin{eqnarray}
\label{bab}
B_{AB} \ &=& \ \left(\nabla^{a}\nabla^{b}+\frac{R_{(1)}}{4}g^{ab}\right)C_{Aa B b }\nonumber \\[2mm]
&+& \ \left(\nabla^{c}\nabla^{d}+\frac{R_{(2)}}{4}g^{cd}\right)C_{Ac B d}\, .
\end{eqnarray}
This can be reduced further by employing~(\ref{Ricci2D}) and (\ref{Riemann2D}). In such a case, the Weyl tensor components read
\begin{eqnarray}
&&C_{Aa Bb  } \ = \ \tfrac{1}{6}\left[R_{(1)} \ + \ R_{(2)}\right]\big(g_{AB}g_{ab} \ - \ g_{Ab}g_{aB}\big)\, , \nonumber \\[2mm]
&&C_{Ac B d}\ = \ -\tfrac{1}{12}\left[R_{(1)} \ + \ R_{(2)}\right]g_{AB}g_{cd}\, . 
\end{eqnarray}
Putting this together we get 
\begin{widetext}
\begin{equation}
  B_{AB} \ = \ -\frac{1}{6}\nabla_A \nabla_B R_{(1)} \ + \   \left[-\frac{1}{12}\nabla^2_{(2)}R_{(2)}\ - \ \frac{1}{24}R_{(2)}^2 \ + \ \frac{1}{6}\nabla^2_{(1)}R_{(1)} \  + \ \frac{1}{24}R_{(1)}^2\right]g_{AB}\, .
  \label{III.35.gh}
\end{equation}
\end{widetext}
An analogous expression holds for the second subspace, obtained by interchanging $(1) \leftrightarrow (2)$ and relabeling the indices $A,B$ as $C,D$

In a similar fashion, we now reorganize the EM stress–energy tensor (\ref{EMt}). In particular, this enables us to express the Lagrange density
$F_{\mu \nu}F^{\mu \nu}$ in the form 
\begin{eqnarray}
F_{\mu \nu}F^{\mu \nu} \ &=& \ 2\left[F_{01}F^{01} \ + \ F_{23}F^{23}\right]\nonumber \\[2mm]
&=& \ 2\left\{(F^{01})^2\left[g_{00}\ \! g_{11} \ - \ (g_{01})^2\right]\right.\nonumber \\[2mm]&&+ \ \left.
 (F^{23})^2\left[g_{22}\ \!g_{33} \ - \ (g_{23})^2\right]\right\}\nonumber \\[2mm]
&=& \ 2\left[(F^{01})^2g_{(1)} \ + \ (F^{23})^2g_{(2)}\right]\, .
\label{Frozdel}
\end{eqnarray}
Here $g_{(1)}$ and $g_{(2)}$, represent the determinants of the metric of the respective 2D subspaces.
The term $F_{\mu \lambda}F_{\nu}{}^{\lambda}$ vanishes whenever $\mu$ and $\nu$ belong to different 2D subspaces. Consequently, in such cases the entire EM stress–energy tensor satisfies $T_{\mu \nu}^{\mathrm{EM}} = 0$. We may therefore restrict our attention only to $T_{AB}^{\mathrm{EM}}$. In this case, one finds that
\begin{eqnarray}
F_{A\lambda}F_{B}{}^{\lambda} \ = \  g_{AB} (F^{01})^{2} g_{(1)} \, ,
\label{Fsmis}
\end{eqnarray}\\[-2mm]
with an analogous expression holding for $T_{CD}^{\mathrm{EM}}$. Combining these results finally gives
\begin{eqnarray}
T_{AB}^{\mathrm{EM}} \ = \  \frac{g_{AB}}{2 e^{2}} \Big[ (F^{01})^{2} g_{(1)} \ - \  (F^{23})^{2} g_{(2)} \Big] \, .
\label{III.38.kl}
\end{eqnarray}
By combining Eqs.~(\ref{III.35.gh}) and~(\ref{III.38.kl}), we arrive at the Bach equation~(\ref{Bachnonvac}) for the first subspace. An analogous derivation applies to the second subspace. The Bach equations can be conveniently expressed in a unified manner (valid for both subspaces). For the $(\alpha)$-part of the metric, they read
\begin{widetext}
    \begin{equation}
-4\nabla_i \nabla_j R_{(\alpha)}+  \left\{-2\nabla^2_{(\beta)}R_{(\beta)}-R_{(\beta)}^2+4\nabla^2_{(\alpha)}R_{(\alpha)}+R_{(\alpha)}^2+ \frac{6G_{_{\!\rm W}}^2\sigma_{(\alpha)}}{e^2} \left[(F^{01})^2g_{(1)}-(F^{23})^2g_{(2)}\right]\right\}g_{ij} \ = \ 0\, ,   
 \label{withtrace}
    \end{equation}
\end{widetext}
where $\sigma_{(\alpha)}$ denotes the signature of the determinant of the $(\alpha)$-part of the metric, and $i,j$ label indices within that sector.
The sub-indices $(\alpha)$ and $(\beta)$ denote distinct 2D subspaces of (\ref{2+2m}) 
[that is, $(\alpha)=(1),  (\beta)=(2)$ or vice versa].

We note in passing that mixed Bach tensor components, such as $B_{AC}$, are identically zero, since the only surviving Weyl tensor components are $C_{AaCc}$, which vanish after applying the mixed covariant derivative $\nabla^a \nabla^c$ [see the explicit form of the Bach tensor in~(\ref{Bach})].

Taking a (2D) traces of (\ref{withtrace}) gives 
\begin{eqnarray}
\label{1tf}
2\{\cdots\} \ - \ 4\nabla^2_{(\alpha)} R_{(\alpha)} \ = \ 0\, , 
\end{eqnarray}
which inserted back to~(\ref{withtrace}) gives
\begin{equation}
\label{tracefree}
\nabla_i \nabla_j R_{(\alpha)} \ = \ \frac{g_{ij}}{2}\nabla^2_{(\alpha)} R_{(\alpha)}\, . 
\end{equation}
By defining  $\partial_a R_{(1)}=\bar \xi_{(1)a}$ and $\partial_c R_{(2)}=\bar \xi_{(2)c}$, Eq.~(\ref{tracefree}) can be equivalently expressed as
\begin{eqnarray}
&&\nabla_a \bar \xi_{(1)b} \ + \ \nabla_b \bar \xi_{(1)a} \ = \ g_{ab} \nabla^i \bar \xi_{(1)i}\, ,\nonumber \\[2mm]
&&\nabla_c \bar \xi_{(2)d} \ + \ \nabla_d \bar \xi_{(2)c} \ = \ g_{cd} \nabla^i \bar \xi_{(2)i}\, ,
\end{eqnarray}
which shows that both $\bar \xi_{(1)a}$ and $\bar \xi_{(2)c}$
are CKVFs 
within their respective 2D subspaces. 
Moreover, by further contracting Eq.~(\ref{tracefree}) with the $(\alpha)$ part 2D mixed Levi–Civita tensor $\epsilon_{k}^{\ \!j}$ and subsequently symmetrizing over the indices $i$ and $k$, we arrive at the KVF equations
\begin{eqnarray}
\nabla_i \xi_{(\alpha)k} \ + \ \nabla_k \xi_{(\alpha)i} \ = \ 0\, ,
\end{eqnarray}
with $\xi_{(\alpha)k}=\epsilon_{k}^{\ \!  m}  \partial_m R_{(\alpha)} = \epsilon_{k}^{\ \!  m} \bar\xi_{(\alpha)m}$ representing KVFs. So each 2D part has one KVF. By further defining $\xi_{(1) \mu}=0$ for $\mu \in \{c, d\}$ and $\xi_{(2) \mu}=0$ for $\mu \in \{a, b\}$ we have 2 independent (and commuting) KVF for the 4D metric.

At this stage, we note that if the vector field $\xi_{(\alpha)}$ is {\em trivial}, then  $R_{(\alpha)}$ must inevitably be constant. In such a case, the $(\alpha)$-subspace is a maximally symmetric 2D subspace with constant Gaussian curvature. As is well known (see the previous section), such spaces admit three KVFs, at least one of which is non-null. 
On the other hand, when $\xi_{(\alpha)}$ is {\em non-trivial} then, as we shall see shortly, it cannot be null. 
Therefore, the preceding reasoning ensures that there are at least two non-trivial, independent, commuting and non-null KVFs in the ST~(\ref{2+2m}). This conclusion will be crucial in the following subsection.

To demonstrate that a non-trivial $\xi_{(\alpha)}$ cannot be null, 
let us assume the contrary --- that $\xi_{(\alpha)}$ is null --- and show that, under this assumption, it must necessarily be trivial. 
Specifically, suppose that $0=\xi_{(\alpha)}^\mu \xi_{(\alpha)\mu}=\xi_{(\alpha)}^k \xi_{(\alpha)k}=-(\nabla_i R_{(\alpha)})(\nabla^i R_{(\alpha)})$, differentiating with respect to $\nabla_j$, and using Eq.~(\ref{tracefree}), we obtain $(\nabla_i R_{(\alpha)})(\nabla^2_{(\alpha)} R_{(\alpha)})=0$. This again implies that $R_{(\alpha)}$ must be constant, either directly if the first factor in the product, i.e. $\nabla_i R_{(\alpha)}$ vanishes, or by virtue of Eq.~(\ref{1tf}): if $\nabla^2_{(\alpha)} R_{(\alpha)}=0$, then $\{\cdots\}=0$. To analyze the relation $\{\cdots\}=0$, we first recall that the Maxwell equations take the form 
$0 = \nabla_\mu F^{\mu \nu} = {\partial_\mu \left( \sqrt{|g|} \, F^{\mu \nu} \right)}/{\sqrt{|g|}}$. 
This implies that the entire EM contribution to $\{\cdots\}$ must be constant, so schematically
\begin{eqnarray}
&&\mbox{\hspace{-10mm}}\{\cdots\} \ = \ 0\nonumber\\[2mm]   &&\Leftrightarrow \;\;\; \;\; 2 \nabla^2_{(\beta)}R^2_{(\beta)} +  R^2_{
(\beta)} \
+ \ const. \nonumber \\[2mm]
&& \mbox{\hspace{5mm}}= \  4 \nabla^2_{(\alpha)}R^2_{(\alpha)}  +   R^2_{
(\alpha)} \ \equiv  \ + R^2_{
(\alpha)}\, .
\end{eqnarray}
Since both equation sides have different regions of definition, they must each be constant. In the last step, we used the fact that $\nabla^2_{(\alpha)} R_{(\alpha)} = 0$. Consequently, $R^2_{(\alpha)}$ is constant, which implies that the null vector $\xi_{(\alpha)}$ is necessarily trivial.

\subsection{Canonical form of metrics and 4D solutions}
\label{reseni}

The key outcome of the previous subsection is that the metrics under consideration always admit at least two non-null, independent, and commuting KVFs. Here, we employ these symmetries to cast the metric into a more convenient form. In particular, we want to demonstrate that each of the 2D metric components can be brought to a diagonal form that is independent of one coordinate. To this end, let us now consider a general 2D space possessing a non-null KVF $\xi$.

We begin with the general form 
\begin{eqnarray}
\xi \ = \  \xi^{m}(x^{m}, x^{n})\,\partial_m \ + \ \xi^{n}(x^{n}, x^{m})\,\partial_n\,  ,
\label{III.B.45.kl}
\end{eqnarray}
where no summation over indices is implied, and $x^{m}, x^{n}$ denote the two coordinates of the 2D subspace.  
Next, we perform a coordinate transformation such that $\xi = \partial_{\psi}$ for some new coordinate $\psi$.  
This is always possible because $\xi$ can be expressed as a linear combination of the basis vectors associated with the 2D KVFs.
Such basis vectors incorporate all symmetries of respective flat 2D-spaces. Specifically, they take the form given in~(\ref{X spatial}) for a Euclidean 2D-space or~(\ref{X lorentzian}) for a Lorentzian 2D-space, both with $K=0$, in appropriate coordinates, where $c_i$ are fixed constants. 

The aforementioned coordinate transformation can then be performed by defining
\begin{eqnarray}
\tilde{x}^{m} \ = \  \int \frac{dx^{m}}{\xi^{m}}\, , \qquad 
\tilde{x}^{n} \ = \  \int \frac{dx^{n}}{\xi^{n}}\, ,
\end{eqnarray}
where $\xi^{m}$ and $\xi^{n}$ depend only on $x^{m}$ and $x^{n}$, respectively. This transformation brings the KVF~(\ref{III.B.45.kl}) to the form
\begin{eqnarray}
\xi \ = \  \partial_{\tilde m} \ + \  \partial_{\tilde n}\, .
\end{eqnarray}
Introducing a new coordinate 
\begin{eqnarray}
\psi \ = \  \tfrac{1}{2}(\tilde{x}^{m} \ + \  \tilde{x}^{n})\, , 
\end{eqnarray}
and choosing the second coordinate, say $\phi$, as $\phi =  (\tilde{x}^{m}  -   \tilde{x}^{n})/2$, we obtain
\begin{eqnarray}
\label{xi jako s}
\xi \ = \  \partial_{\psi}\, .
\end{eqnarray}
Since $\xi$ is KVF, the metric can be expressed independently of $\psi$ as
\begin{eqnarray}
\label{2Dmphi}
ds^2 \ =  g_{\psi\psi}(\phi)\, d\psi^2  + \ 2 g_{\psi\phi}(\phi)\, d\psi\, d\phi + \  g_{\phi\phi}(\phi)\, d\phi^2.
\end{eqnarray}
Let us now diagonalize this metric. In two dimensions, any metric is locally diagonalizable. However, the key question is whether it can be diagonalized in such a way that the independence from one of the coordinates is preserved. 

To explore this, let us keep $\phi$ as one of the coordinates and define a new coordinate, say $\chi$, as follows
\begin{eqnarray}
\chi \ = \  2\psi \ + \  \int \frac{g_{\psi \phi}(\phi)}{g_{\psi \psi}(\phi)} \, d\phi\, .
\end{eqnarray}
It is easy to see that this transformation  renders the metric diagonal, and the components remain independent of $\chi$.
However, the transformation breaks down when $g_{\psi \psi} = 0$. If this holds globally,  the Ricci scalar (or equivalently the Riemann tensor) is zero, and thus the space is flat. Since any two-dimensional flat space can be brought into a diagonal form via an appropriate coordinate transformation, a different diagonalization is still possible. In case the problem is only local, the transformation must be defined differently  in the neighborhood of the points where $g_{\psi \psi}(\phi) = 0$.

Once we have metric in the form
\begin{eqnarray}
ds^{2}\ =\ g_{\chi \chi}(\phi)d\chi^{2}+g_{\phi \phi}(\phi)d\phi^{2} \, ,
\end{eqnarray}
the ``canonical'' form can be obtained by multiplying the second term by unity, expressed as $ \sigma_{(\alpha)} g_{\chi \chi}/(\sigma_{(\alpha)} g_{\chi \chi} )$, and introducing a new coordinate 
\begin{eqnarray}
\vartheta \ = \ \int \sqrt{\vert g_{\chi \chi} g_{\psi \psi}\vert}\ \! d\phi \, ,
\end{eqnarray}
which transforms the metric [when applied to the $(\alpha)$ sector of (\ref{2+2m})] into
\begin{equation}
\label{canonical}
ds^{2}_{(\alpha)} \ = \ \sigma_{(\alpha)} B(\vartheta)d\chi^{2}\ + \ B^{-1}(\vartheta)d\vartheta^{2}\, ,
\end{equation}
where $B(\vartheta) $ is $\sigma_{(\alpha)}g_{\chi \chi} $ written as a function of the new coordinate $\vartheta$.  

Now, let us determine a specific functional form of $B(\vartheta)$. From~(\ref{tracefree}) and using the identity for the 
Laplace--Beltrami operator 
\begin{equation}
\nabla^{2} f \ = \  \frac{\partial_{\mu}\!\left(\sqrt{\lvert g \rvert}\, g^{\mu\nu} \partial_{\nu} f \right)}{\sqrt{\lvert g \rvert}}\, ,
\end{equation}
we obtain
\begin{equation}
R_{(\alpha)} \ = \  -B''(\vartheta)\, ,
\end{equation}
Substituting this into the $\vartheta\vartheta$-component of Eq.~(\ref{tracefree}) yields
\begin{equation}
\label{B4polynom}
\frac{d^{4} B(\vartheta)}{d\vartheta^{4}} \ = \  0\, ,
\end{equation}
which implies that $B(\vartheta)$ is a cubic polynomial. The remaining trace-free equations provide no additional constraints. The same procedure should be applied to both sectors of the metric, after which the results should be combined to create the complete 4D metric.

At this point, we employ the fact that both components of the EM field are constant. As mentioned in Sec.~\ref{reseni}, this follows directly from the Maxwell equations, $\nabla_\mu F^{\mu\nu} = 0$. We denote these constants corresponding to the electric and magnetic charges as
$q_1 = F_{01}$ and $q_2 = F_{23}$,  respectively.

The algebraic relations among the constants associated with both 2D metric sectors and the EM field are provided by Eqs.~(\ref{withtrace}), yielding complete solutions of the form
\begin{eqnarray}
ds^2  &=&   -A(r)\, dt^2 \ + \  A^{-1}(r)\, dr^2 \nonumber \\[2mm] 
&&+ \  F^{-1}(x)\, dx^2 + F(x)\, dy^2\, ,
\label{III.58.kl}  
\end{eqnarray}
with
\begin{eqnarray}
\label{resenikomplet}
&&A(r) \ = \   a_0 \ + \  a_1 r \ + \  a_2 r^2 \ + \  a_3 r^3\, , \nonumber \\[2mm]
&&F(x) \ = \  f_0 \ + \  f_1 x \ + \  f_2 x^2 \ + \  f_3 x^3\, ,
\end{eqnarray}
and the additional algebraic constraint
\begin{equation}
3 a_1 a_3 \ - \  3 f_1 f_3 \ + \  f_2^2 \ - \  a_2^2 \ = \  \frac{3 G_{_{\!\rm W}}^2}{2 e^2} \left( q_1^2 \ + \  q_2^2 \right).
\label{resenikompletAP}  
\end{equation}  
For a detailed discussion of this solution, see Sec.~\ref{properties}. 
It is worth emphasizing that, although its derivation involved a careful analysis of the field equations and their traces,
once the canonical form (\ref{canonical}) is established for both metric sectors, solving the full 4D Bach equations becomes straightforward.

Before proceeding, it is necessary to clarify a few points. Since both $A(r)$ and $F(x)$ are third-order polynomials, they can generally change sign 
(full third order polynomials have at least one real root). When crossing a root of $F(x)$, the metric signature changes from $(1,3)$ to $(3,1)$ (or vice versa). Although the metric remains Lorentzian, this change violates Sylvester's law of inertia~\cite{Norman}. Consequently, we must restrict ourselves to coordinate ranges in which $F(x)$ is positive when working in this coordinate system.

For $A(r)$, Sylvester's law of inertia is not violated when crossing roots, since only the signs of two coordinates are exchanged and the overall metric signature remains unchanged. These roots --- up to three in total --- correspond to the Killing horizons associated with the KVF $\xi_{(1)} = \partial_t$, where the roles of $r$ and $t$ exchange their meaning. The coordinate $\vartheta$ has been relabeled as $r$, which might suggest that it is automatically spacelike at infinity, however, this is not necessarily the case. The class of solutions includes both configurations where the asymptotic coordinate is spacelike and those where it is timelike.

Let us conclude with a brief remark on the assumption regarding the EM field introduced in Eq.~(\ref{polesep}), expanding upon the discussion presented in Sec.~\ref{hlavni cast}. Although the assumption~(\ref{polesep}) is natural in the context of the product metric~(\ref{2+2m}), as it ensures ``symmetry sharing'', it is worthwhile to consider possible generalizations. In particular, when we allow the nonzero components $F_{01}$ and $F_{23}$ to be general functions of the coordinates, we obtain the same formal results: Eq.~(\ref{withtrace}) retains its structure, as does the procedure involving the KVFs and the canonical rearrangement~(\ref{canonical}) of the 2D metrics.  Applying Maxwell's equations in this generalized setting restricts the functional dependence to $F_{01}(x^2, x^3)$ and $F_{23}(x^0, x^1)$. However, upon substituting these back into Eq.~(\ref{withtrace}), we find that both components must, in fact, be constant. Hence, the generalized assumption is equivalent in practice to the original one. 
A fully general EM field, however, would not be compatible with the present framework: in that case, an additional term of the form $g^{ab} g^{cd} F_{ac} F_{bd}$ would appear in Eq.~(\ref{Frozdel}), and after varying~(\ref{polesep}), it would no longer be possible to express the resulting equations in a form consistent with the existence of KVFs.

\subsection{Constant-curvature-part cases: relation to BRT}
\label{Rcomment}

The case in which one of the 2D sectors of the metric possesses constant Gaussian curvature can be treated either as a special sub-case of our previous analysis or, alternatively, following Riegert's approach \cite{Riegert}, in which the two-dimensional action is obtained from Eq.~(\ref{akce}) by integrating over the constant-curvature sector, and the corresponding field equations are then derived from this reduced action. As we have seen, when one of the two-dimensional subspaces~$(\alpha)$ has constant curvature --- either Euclidean~(\ref{2DE}) or Lorentzian~(\ref{2DL}) --- the associated vector field~$\xi_{(\alpha)}$ becomes trivial.
Nevertheless, other KVFs, given by~(\ref{X spatial}) or~(\ref{X lorentzian}), can play an analogous role. These cases are characterized by a constant Ricci scalar, $R_{(\alpha)} = 2 K_{(\alpha)}$, where $K_{(\alpha)}$ is the corresponding Gaussian curvature. Imposing symmetry on the EM field --- that is, requiring that the Lie derivative  of the EM potential $A_\mu$ with respect to the KVFs~(\ref{X spatial}) or~(\ref{X lorentzian}), i.e. $\mathcal{L}_X A_\mu = 0$, vanishes --- immediately yields an EM potential one-form
$\mathcal{A} = A_k(x^m, x^n) \, dx^k$,
where $x^m, x^n$ are the coordinates of the $(\beta)$ sector, and the summation extends only over this part.

In passing, we note that, as will be seen in Sec.~\ref{zaklad}, the constant-curvature cases correspond to $a_3$ (or $f_3$) in Eq.~(\ref{resenikomplet}) being zero, with $K_{(1)}=-a_2$ or $K_{(2)}=-f_2$. It can be observed that if the corresponding symmetry (i.e. spherical, planar, or hyperbolical) is not imposed on the EM field, the solution may possess both $q_1$ and $q_2$. However, since we aim to comment on Riegert's approach here, we assume that the symmetry is imposed, so that at most one of the charges can be present.

Apart from the derivation of the field equations, we have, in the preceding subsections, applied a generalized form of Riegert's procedure, leading up to the derivation of the KVFs and establishing their non-null character. However, in the subsequent parts of his proof, various issues arise.  Riegert first asserts that, for a timelike KVF $\xi$, the metric can be diagonalized and $\xi$ can be made ``explicitly timelike'' [Eqs.~(19a) and (19b) in his paper]. However, this step is neither commented on nor explained; for this reason, we have included the explicit construction in Sec.~\ref{reseni}. Using the fact that $\xi$ is a KVF, he then shows that this form is, in fact, the final form required, independent of the timelike coordinate [Eq.~(21) in his paper]. 

Riegert further asserts that since two coordinates of the Lorentzian metric sector are interchangeable, all metrics can be brought into a form independent of the timelike coordinate. 
Here, however, a subtlety arises. For a two-dimensional Lorentzian metric with signature $(1,1)$, the distinction between ``time'' and ``space'' is essentially conventional; either coordinate can be designated as timelike, so the interchangeability holds. 
However, if we wish to make statements about the 4D metric, of which this 2D metric is a part, the situation changes. A 4D Lorentzian metric must have signature $(1,3)$ or $(3,1)$, so the ``time'' coordinate is necessarily the one with a sign opposite to that of the remaining coordinates (i.e., in a diagonal metric, it is the coordinate whose squared differential is multiplied by a function of opposite sign). Additionally, Riegert implicitly assumes that the KVF does not change sign, and the metrics he presents cannot interchange the character of the two coordinates.  
The resulting metric [Eq.~(21) in Riegert's paper] is static everywhere, which, as we have already seen, is not the case even for the MK solution~(\ref{MK}). [Although MK is presented as a unique solution, Riegert's metric~(21) does not actually encompass all its versions and sectors.] 
Consequently, the final solution classes obtained from Riegert's procedure, using his notation, should consist of two distinct classes rather than a single one, and the interchange of coordinate characters should be permitted. We can thus conclude that, in cases where one 2D sector has constant curvature --- adopting the coordinates introduced in Sec.~\ref{const_curv}, which are generally more convenient when working with these two-spaces --- the resulting forms are
\begin{eqnarray}
\label{RDE}
ds^{2} =  -B(r)dt^{2} \ + \ B^{-1}(r)dr^{2} \ + \ \frac{2d\zeta d\bar{\zeta}}{\left(1+\frac{K\zeta \bar{\zeta}}{2}\right)^{2}}\, ,~~~
\end{eqnarray}
and
\begin{eqnarray}
\label{RDL}
ds^{2}  =  -\frac{2du dv}{\left(1-\frac{K u v}{2}\right)^{2}} \ + \ B^{-1}(x)dx^{2} \ + \ B(x)dy^{2}\, ,
\end{eqnarray}
with 
\begin{eqnarray}
B(\vartheta) \ = \ b_3 \vartheta^{3} \ + \ b_2 \vartheta^{2} \ + \ b_1 \vartheta \ + \ b_0\, ,
\end{eqnarray}
and
\begin{equation}
\label{tridy}
K^2 \ + \  3b_1b_3 \ - \  b_2^2 \ = \ \frac{3 \sigma G_{_{\!\rm W}}^2 q^2}{2e^2} \ = \ 0\, .
\end{equation}
The parameter $\sigma$ corresponds to the sign of the determinant of the constant-curvature sector, and $q$ represents the single component of the EM field allowed by the symmetry. The first class corresponds to Riegert's solution for $K=1$ [and positive $B(r)$]. We again emphasize that $A(r)$ can be either positive or negative at infinity, meaning that $r$ may be spacelike or timelike in that region; consequently, the first metric actually encompasses two distinct classes. The function $B(x)$ must be strictly positive.

The aforementioned issue is not the only problem encountered in Riegert's proof. Another concern is that Riegert neglects the possibility of singular Weyl transformations, which implies that the STs constructed from his solutions are generally only locally equivalent and may exhibit differences in their causal or global structure. We address this point in Sec.~\ref{tridyE}.

\subsection{Yang--Mills fields}
\label{YM}

Here, we extend the results obtained for the EM case to more general, non-abelian YM fields. The extension is relatively straightforward. To this end, we assume that the resulting field strength (technically, the curvature 2-form) takes a separated form that is formally analogous to~(\ref{polesep}), but now with the following  modifications:
\begin{eqnarray}
\label{YMAF}
&&F_{\mu \nu} \ \mapsto \ {\mathbf{F}}_{\mu \nu} \  = \ F_{\mu \nu}^q \ \! {\mathbf{T}}_q\, , \nonumber \\[2mm] 
&&A_\mu \ \mapsto \ {\mathbf{A}}_{\mu}\ =A_\mu^q \ \! {\mathbf{T}}_q   \, .
\end{eqnarray}
Here, ${\mathbf{T}}_q$ are matrix-valued generators of the Lie algebra, with the index $q$ running from $1$ to the dimension of the ensuing Lie group. The quantities $F_{\mu \nu}^q$ and $A_\mu^q$ are real-valued functions representing the field strengths and gauge potentials, respectively. The relationship between the (matrix-valued) field strength and the (matrix-valued) gauge potential is given by
\begin{equation}
{\mathbf{F}}_{\mu \nu} \ = \ \partial_\mu {\mathbf{A}}_\nu \ - \  \partial_\nu {\mathbf{A}}_\mu \ - \ i[{\mathbf{A}}_\mu, {\mathbf{A}}_\nu]\, ,
\end{equation}
and the generators satisfy
\begin{eqnarray}
\label{generators}
&&\mathrm{Tr}({\mathbf{T}}_q{\mathbf{T}}_p) \ = \ k\delta_{qp}\, , \;\;\;\quad k \ > \  0\, , \nonumber \\[2mm]
&&[{\mathbf{T}}_q, {\mathbf{T}}_p] \ = \ if_{qps}{\mathbf{T}}_s,
\end{eqnarray}
where $f$'s are structure constants. The positivity of $k$ is always guaranteed for simple compact groups. For the fundamental representation of 
$SU(N)$, which is common in particle physics, $k$ is typically set to $1/2$ (see, e.g.,~\cite{Rubakov}).

The stress energy tensor for YM fields reads 
\begin{equation}
T_{\mu \nu} \ = \ \frac{1}{\textsl{g}^{2}k}\mathrm{Tr} \left( {\mathbf{F}}_{\mu \lambda}{\mathbf{F}}_{\nu}^{\: \:\lambda} \ - \ \frac{1}{4}g_{\mu \nu}{\mathbf{F}}_{\alpha \beta}{\mathbf{F}}^{\alpha \beta} \right)\, ,
\end{equation}
where $\textsl{g}^{2}$ is a coupling constant. Field equations (i.e., analogs of Maxwell equations) read
\begin{equation}
\label{YMfe}
D_\mu {\mathbf{F}}^{\mu \nu}  \ \equiv \   \nabla_\mu {\mathbf{F}}^{\mu \nu}-i[{\mathbf{A}}_\mu, {\mathbf{F}}^{\mu \nu}] \ = \ 0\, .
\end{equation}
Here $D_{\mu}$ is the YM covariant derivative.
Thanks  to normalization of the Lie algebra generators [trace relation~(\ref{generators})], the stress energy tensor can be rewritten in terms of field strength components as
\begin{equation}
T_{\mu \nu} \ = \ \frac{1}{\textsl{g}^{2}} \sum_q \left( F^{q}_{\mu \lambda}F_{\nu}^{q\: \:\lambda} \ - \ \frac{1}{4}g_{\mu \nu} F^{q}_{\alpha \beta}F^{q \: \alpha \beta}\right)\, .
\end{equation}
Consequently, in our case, if we assume the separation condition~\eqref{polesep} for all 
$\mathcal{F}^q$ (the most natural extension of the EM case), then 
Eqs.~\eqref{Frozdel} and~\eqref{Fsmis} must hold for each $q$, by the same arguments as in the EM case. We can thus write  for the $(\alpha)$ part of the metric
\begin{eqnarray}
\label{YMTnonz}
\mbox{\hspace{-2mm}}T_{ij}
=  - \frac{\sigma_{(\alpha)}g_{ij}}{2\textsl{g}^{2}}\sum_q \left[(F^{q \:01})^{2}g_{(1)} \ - \ (F^{q \: 23})^{2}g_{(2)} \right]\, ,
\end{eqnarray}
where $i,j$ are the $(\alpha)$ part indices --- again, the mixed components must be zero. Eq.~(\ref{YMTnonz}) is a clear analogue of Eq.~(\ref{III.38.kl}). 

As in the (electro-)vacuum case, the Bach equations can be written in the same formal form as Eq.~(\ref{withtrace}), namely
\begin{eqnarray}
-4\nabla_i \nabla_j R_{(\alpha)} \ + \ g_{ij} \{\ldots\} \ = \ 0\, ,
\end{eqnarray}
implying that the ST admits the same two KVFs as in those cases. We now aim to demonstrate that they cannot be simultaneously null and nontrivial. To this end, let us assume the opposite. In that case, one can proceed from the general metric form ~(\ref{III.B.45.kl}) to the form (\ref{xi jako s})-(\ref{2Dmphi}). Since $\xi$ is required to be null, $g_{\chi\chi}$ must vanish, which implies a flat ST (and consequently a trivial $R_{(\alpha)}$ and KVF), as already discussed in the EM case. Based on this reasoning, we conclude --- analogously to the EM case --- that there always exists at least one nontrivial, non-null KVF for each 2D metric sector. Hence, its canonical form [Eq.~(\ref{canonical}) together with~(\ref{B4polynom})] remains the same. 
Therefore, the metric structure (\ref{III.58.kl})-(\ref{resenikompletAP}) remains unchanged, with the only modifications arising from the coupling constant on the RHS of~(\ref{resenikompletAP}) and from the fact that the quantities $q_1^2$ and $q_2^2$ are now given by the sums of $(F^{q\,01})^2$ and $(F^{q\,23})^2$, respectively, as specified in~\eqref{YMTnonz}. Both sums must be constants [again, by Eq.~(\ref{withtrace})], but the particular $F^{q\,01}$ and $F^{q\,23}$ do not need to be.

To construct and analyze explicit solutions, one would need to specify a particular Lie algebra, which would in turn determine the form of the YM equations. Nevertheless, independent of the particular algebra, there always exists a simple class of solutions. Assuming that each component $F^{q\,01}$ and $F^{q\,23}$ is itself constant, we may take
\begin{eqnarray}
\label{YMvolbaA}
&&{\mathbf{A}}_0 \ = \ -\int {\mathbf{F}}_{01} \ dx^{1}\, , \;\;\;  {\mathbf{A}}_1 \ = \ 0\, ,  \nonumber \\[2mm]
&&{\mathbf{A}}_2 \ = \ 0\, , \;\;\; {\mathbf{A}}_3 \ = \ \int {\mathbf{F}}_{23} \  dx^{2}\, ,    
\label{D.74.kl}
\end{eqnarray}
which automatically satisfies the YM field equations~(\ref{YMfe}), independently of the structure constants --- or, equivalently,  of the particular Lie algebra considered. So, there are no additional constraints on the constants $F^q_{01}$ and $F^q_{23}$ in this solution, there are $2d$  ($d$ is the dimension of the Lie algebra), arbitrary constant squares on the right-hand side of condition~(\ref{resenikompletAP}).

\subsection{Local equivalence of STs, equivalence classes and subclasses}
\label{tridyE}

In Sec.~\ref{reseni}, we obtained all 2$+$2 direct-product metrics~(\ref{2+2m}) that solve the Bach equations~(\ref{Bachnonvac}) in the presence of an EM field and are compatible with the structure~(\ref{polesep}). A corresponding generalization to the YM fields, preserving the same metric structure, was discussed in Sec.~\ref{YM}.
We have seen in Sec.~\ref{WCG} that it is not possible to change the symmetry (i.e. presence of CKVFs or CKTs) by Weyl transformation. Accordingly, we can say the WCG is a theory of Weyl-equivalent classes of STs, where all STs in one class share CKVFs and CKTs (and, more generally, any symmetries invariant under Weyl rescaling).

However, as shown in Sec.~\ref{WCG}, degenerate Weyl transformation (i.e., when $\Omega$ becomes singular or vanishes) can significantly alter both the global topological and causal properties of a ST. Consequently, all STs generated from those obtained here by degenerate $\Omega$ should be regarded only as \emph{locally} equivalent, whereas equivalence under a regular $\Omega$ is \emph{global}. This suggests the existence of finer equivalence subclasses that account for these properties. Shifts between different subclasses are possible only via irregular transformations, while within the same subclass, transformations can be performed using regular $\Omega$.

Although general methods for classifying global ST structures exist (see, e.g., Ref.~\cite{Malament} and citations therein), our goal here is to identify a more direct computational criterion. We therefore focus on the quantities that can be affected by a degenerate transformation --- but remain unchanged under a non-degenerate one --- particularly for the STs considered in this work. (The underlying ideas, however, are quite general.) While there are certainly many such properties that could, in principle, be computed and used to compare different STs, we have already seen in Sec.~\ref{WCG} that any zero or divergence in $\Omega$ is reflected in $\Psi_i$ and in invariants such as~(\ref{ID}). Owing to their universality, we shall particularly focus on these structures. The corresponding changes are associated with the emergence, cancellation, or amplification of scalar polynomial curvature singularities and emergence of the ``$O$-type'' parts of the ST (i.e., portions of the ST where the Weyl tensor vanishes and the geometry is therefore conformally flat). We also note the possible modifications of (conformal) Killing horizons, which correspond to the zeros of the (C)KVF norm. Detailed calculations will be presented in Sec.~\ref{zaklad}. Here we only remark that all solutions~(\ref{resenikomplet}) are of Petrov type~$D$ (or~$O$), with the invariant (\ref{ID}) being
\begin{eqnarray}
\label{I}
    I &=& 3\Psi_2^2 \ = \ \frac{[A''(r) \ + \ F(x)'']^2}{48}\nonumber \\[2mm]
    &=& \ \frac{(a_2 \ + \ 3a_3r \ + \ 3f_3x \ + \ f_2)^2}{12}\, .
\end{eqnarray}
In the following, we denote $A''(r) + F''(x) \equiv h(r, x)$.

We observe that a degenerate transformation can, in principle, generate scalar polynomial curvature singularities of virtually any type --- featuring an arbitrary number of singularities with arbitrary dimensionality, ST character, position relative to horizons, and ``strength''. Some of these properties, such as the number, dimensionality, and general ``shape'', can be characterized by homotopy groups. However, such analysis provides little insight into more ``qualitative'' features, such as the ST character or the position relative to (conformal) Killing horizons, which must be determined separately. The formation of a singularity is associated with zeros of the Weyl factor $\Omega$: since $\tilde{I} = I\,\Omega^{-4}$, a singularity arises at points where $\Omega$ vanishes, provided that the corresponding zero in the denominator of $\tilde{I}$ is not canceled by a zero in the numerator. If a singularity already exists at that location, it is merely amplified.
The region of the newly formed singularity generally also corresponds to a zero of the (C)KVF norm --- formally a (conformal) Killing horizon --- for all (C)KVFs considered. Since $\tilde{V}^\mu \tilde{V}_\mu = \Omega^{2} V^\mu V_\mu$, the zeros of $\Omega$ become zeros of the norm, provided they do not cancel an existing divergence or merely reinforce an already existing horizon.

Moreover, zeros or zero-point regions of arbitrary shape can be created (or removed) in individual $\Psi_i$ components and, consequently, in the invariant $I$. Such zero point regions indicate regions of the ST that are conformally flat, i.e., where the Weyl tensor vanishes. These regions are typically generated by divergences in the Weyl factor $\Omega$ (or, equivalently, by zeros in $\Omega^{-1}$). Their characteristics are analogous to those described for the singularity case. A new $O$-type part of the ST arises provided that a zero in $\Omega^{-1}$ is not canceled by a preexisting zero in $I$, or does not merely amplify an already vanishing part of $I$. Divergences in $\Omega$ can simultaneously modify the zeros of the (C)KVF norm, potentially removing a horizon or driving the norm to infinity at the location of the divergence.

In passing, we stress that the cancellation of a (C)KVF horizon should be understood as the disappearance of the region where its norm is zero. However, if the norm originally changed sign when crossing the root, this change must persist, since $\Omega^{2}$ is strictly positive. Consequently, a discontinuity in the norm appears. A similar consideration applies to horizon formation: the zero norm area can be created, but the sign change cannot. 

Analogously to curvature singularities, regions of zero norm of the (C)KVF can exhibit a wide range of characteristics, including the number of the zero-norm regions, their dimensionality, general shape, and position relative to singularities and to one another. Since the existence of such regions represents a global and causal property of the ST, their characteristics must likewise be analyzed and compared across different STs. These features can also be investigated using topological methods --- for instance, by examining the topology of the hypersurfaces defined by the vanishing of the (C)KVF norm, classifying their connectivity through homotopy or homology groups.

We may regard the creation or removal of a scalar curvature singularity, or of a conformally flat region of the ST, as a modification of its global or causal structure. A natural question then arises: if the Weyl factor $\Omega$ is singular but merely amplifies or attenuates existing zeros or divergences in the invariant $I$, does this also constitute a global or causal change? To explore this, let us consider a specific example. Suppose that $I$ has the form given in Eq.~(\ref{I}). It already possesses a zero at $h(r, x) = 0$ (the $O$-type parts) and diverges in the limit $h(r, x) \rightarrow \pm \infty$ (singularities). 
Using, for example, $\Omega^{4} = |h(r, x)|$ not only weakens the existing zero in $I$ but also generates a KVF horizon, i.e. a surface where the KVF norm vanishes, at $|h(r, x)| = 0$. In contrast, when $h(r, x)$ already represents a nondegenerate KVF horizon [which may occur for specific parameter choices in Eq.~(\ref{resenikomplet})], applying $\Omega^{4} = |h(r, x)|^{-1}$ strengthens the existing zero in $I$ but simultaneously eliminates the horizon. Thus, there appear to be cases in which such transformations alter the global structure of the ST.

Two further remarks are in order concerning the limitations of the above approach.
First, it is necessary to determine whether the chosen coordinate chart extends to a global chart on the entire ST manifold.
If it does not (or if this is unknown), inequivalent regions may still
exist in the uncovered portions of the two STs, even when the conformal
factor $\Omega$ relating them appears regular. 
Second, a different strategy is required when studying STs that already
belong to the type~$O$ (which in our case means that $f_{3} = a_{3} = 0$ and $a_{2} = -f_{2}$). In such
cases the invariant $I=0$, and one must rely on alternative techniques --- for example, demonstrating geodesic incompleteness --- to establish the existence of
a singularity. (However, not all singularities need to be reachable by geodesics.)

In the preceding discussion, we examined which features of a ST may change
under a degenerate Weyl transformation, i.e. when the Weyl conformal factor becomes singular or zero. 
These considerations are useful when constructing
one ST from another by applying the Weyl transformation. However, when instead we are
 given two generic STs and aim to determine whether they are Weyl
equivalent, the natural first step is an analysis of their symmetries (CKVFs, CKTs,
etc.). If the two geometries do not share the same symmetry structure, they cannot
belong to the same Weyl class.  If they do, and an explicit transformation relating
them can be identified, the (ir)regularity of that transformation characterizes the
type of equivalence. In practice, however, finding such a transformation --- typically
involving both a Weyl rescaling and a coordinate transformation --- may be nontrivial.
In such cases, one must instead analyze geometric features such as singularities,
regions of the $O$-type, and horizon structure (and possibly additional properties).
Explicit examples illustrating how the invariant $I$ may differ under degenerate transformations, together with their interpretation for the solutions considered
here, will be presented in Secs.~\ref{GRrel} and \ref{WCGrel}.

\section{Basic properties and relation to other solutions}
\label{properties}

Let us now examine several basic properties of the solutions obtained in
Secs.~\ref{reseni} and~\ref{YM}, together with their relations to other geometries
within the same symmetry classes. Our goal here is not to provide a detailed analysis of any specific solution, but
rather to offer a set of remarks that clarify how these solutions fit within the
corresponding equivalence classes and to connect them with existing
results. For convenience, these comments are presented in a numbered list.

\subsection{Basic properties}
\label{zaklad}

\textbf{1)} In Sec.~\ref{hlavni cast} we analyzed the symmetries of 2$+$2 direct-product STs and found that they always admit two second-order KTs and two
KVFs. These four symmetry generators are manifestly independent. Consequently, the geodesic motion is integrable, with integrals of motion 
\begin{eqnarray}
&&I_{1} \ = \ g_{ab}\ \!p^{a}p^{b}\, ,\;\;\;\;  I_{2} \ = \ g_{cd}\ \!p^{c}p^{d}\, ,  \nonumber \\[2mm] 
&&I_{3}\ = \ p^{t}\, , \;\;\;I_{4} \ = \ p^{y}\, .
\end{eqnarray}
Supplementing this with the remaining momenta
\begin{eqnarray}
    p^{r} \ = \ \pm\frac{\sqrt{A(r)I_{1}+I_{3}^{2}}}{A(r)}\, , 
    \;\;\
    p^{x} \ = \ \pm\frac{\sqrt{F(x)I_{2}-I_{4}^{2}}}{F(x)}\, ,~~
\end{eqnarray}
reduces the geodesic equations (in diagonal metric form, with no summation implied)
\begin{eqnarray}
    p_{\mu} \  = \ g_{\mu \mu}\,\frac{dx^{\mu}}{d\lambda}\, ,
\end{eqnarray}
to quadratures. The system is therefore integrable (and in the special cases, when additional independent symmetries are involved, can be even superintegrable, i.e., solvable, at least partially, algebraically). Considering the whole conformal class (where KVFs/KTs generally become CKVFs/CKTs), integrability statement only applies to null geodesics.

Since the EM field~(\ref{polesep}) was considered to be compatible with 
the 2$+$2 product structure (and thus with the KT symmetries), and is moreover constant so that it also respects the KVFs, the charged case proceeds analogously for electro-geodesic motion. The only modification is that the  uncharged Hamiltonian must be replaced by its charged counterpart, in which the 
canonical momentum no longer coincides with the physical (kinetic) momentum $p_\mu$. 
Instead, one must apply the minimal-coupling prescription $p_\mu \longrightarrow \Pi_\mu  - Q A_\mu $, where $Q$ is the particle's charge and $\Pi_\mu$ denotes the generalized (canonical) momentum 
\cite{BHsym}. The Hamiltonian then takes the familiar form
\begin{eqnarray}
    H \ = \ g^{\mu\nu} \, \big(\Pi_\mu - Q A_\mu\big)\big(\Pi_\nu - Q A_\nu\big) \, .
\end{eqnarray}
This Hamiltonian still separates into two independent parts, yielding 
two corresponding integrals of motion. Geodesic motion --- charged or uncharged --- is relevant not only for test-particle dynamics but also, for example, in the study of singularities (such as geodesic incompleteness) and other geometric features.

\vspace{3mm}

\textbf{2)} For the Lorentzian sector of the metric~(\ref{III.58.kl})-(\ref{resenikompletAP}), 
one may introduce generalized Eddington--Finkelstein (EF) coordinates~\cite{EF},  
by defining the tortoise-like coordinate 
\begin{eqnarray}
r^{\ast} \ = \  \int A(r)\, dr \, ,
\end{eqnarray}
and the null coordinate 
\begin{eqnarray}
w \ = \  t \ + \  c\, r^{\ast}\, ,
\label{81.gh}
\end{eqnarray}
(with $c = +1$ for ingoing rays and $c = -1$ for outgoing rays). The ensuing metric, when combined with 
the constant-curvature Euclidean part in Eq.~(\ref{2DE}), takes the form
\begin{eqnarray}
    ds^{2} \ = \  -A(r)\, dw^{2} \ + \  2c\, dw\, dr \ + \  d\Sigma_E^{2}\, ,
\end{eqnarray}
which is a member of the well-known Kundt class of STs~\cite{kniha}. Generalized EF coordinates are also significantly more convenient for many computations (e.g., the Bach equations in 4D) and better suited for analyzing geometric properties, since structures such as horizons are manifestly regular in them.


\vspace{3mm}

\textbf{3)} By choosing the null tetrad as 
\begin{eqnarray}
&&l \ = \  \frac{1}{\sqrt{2A(r)}} \,\partial_t \  + \  \sqrt{\frac{A(r)}{2}}\, \partial_r\, , \nonumber \\[2mm] 
&&n \ = \  \frac{1}{\sqrt{2A(r)}} \,\partial_t \ - \  \sqrt{\frac{A(r)}{2}}\, \partial_r\, , \nonumber \\[2mm] 
&&m  \ = \  \sqrt{\frac{F(x)}{2}}\, \partial_x \ + \  \frac{i}{\sqrt{2F(x)}}\, \partial_y\, ,
\end{eqnarray}
one can find that the only non-vanishing Weyl and Ricci scalars are
\begin{eqnarray}
&&\Psi_2 \ = \  -\frac{1}{12}\,\big[A''(r) \ + \ F''(x)\big]\, , \nonumber \\[2mm]
&&\Phi_{11} \ = \  \frac{1}{8}\,\big[A''(r) \ - \ F''(x)\big]\, ,
\end{eqnarray}
These should be supplemented with the invariant $I$ defined by~(\ref{ID}), which in the present context is given by Eq.~(\ref{I}).  

The latter implies that the solutions are of Petrov type $D$, or type $O$ if $h(r,x)=0$. The latter can occur globally when $a_3 = f_3 = 0$ and $-K_L \equiv a_2 = -f_2 \equiv K_E$, corresponding to the special constant-curvature cases discussed in Sec.~\ref{const_curv}. The local conformally flat portions of Petrov type $D$ STs are determined by the relation between $r$ and $x$ implied by the condition $h(r,x)=0$. The invariant $I$ diverges (indicating scalar curvature singularities) whenever $h(r,x)  \to \pm \infty$.

\vspace{3mm}


\textbf{4)} Since the class of STs under consideration contains many black (and white) hole geometries (see Secs.~\ref{WCGrel} and \ref{GRrel} for further details), it is natural to examine the (conformal) Killing horizons associated with 
$\xi_{(1)} = \partial_t$ from this perspective as well.  In the black-hole case, one of these horizons must coincide with an event horizon, and its location is preserved under Weyl transformations. Moreover, there exists a Weyl invariant definition of surface gravity~\cite{CKVtermo, Ctemper}
\begin{equation}
\nabla_\mu (\xi^\nu \xi_\nu) \ = \  -2\kappa\,\xi_\mu \, ,
\end{equation}
implying the invariant temperature $T=\kappa/(2\pi)$, except in cases where $\partial_\mu \Omega^2|_H$ is singular. For the STs considered here, this yields
\begin{equation}
\kappa \ = \  \frac{c\,A'(r)}{2}\Big|_H\,  .
\end{equation}
The location of the horizon 
$H$ follows from solving the cubic equation $A(r)=0$. An interesting direction for further study is the effect of the vanishing --- or cancellation --- of the (C)KVF on the horizon. This situation corresponds to a singular derivative of $\Omega$ and effectively replaces the horizon with a conformally flat region, although its role as a geometric ``boundary'' must remain intact.



\subsection{Relation to GR solutions}
\label{GRrel}

\textbf{1)} We now examine whether any of our solutions also satisfy the Einstein field equations,
\begin{equation}
G_{\mu\nu} \ + \  \Lambda g_{\mu\nu} \ = \  8\pi G\, T_{\mu\nu}\, ,
\label{IV.87.kl}
\end{equation}
with $G$ denoting Newton's gravitational constant. The application of metric~(\ref{III.58.kl})-(\ref{resenikompletAP}) to~(\ref{IV.87.kl})
immediately yields $a_{3}=f_{3}=0$, indicating that both two-dimensional sectors must have constant Gaussian curvature. The remaining algebraic constraints imposed by (\ref{IV.87.kl}) are
\begin{eqnarray}
\label{Ealg}
&&f_{2} \ + \ \Lambda \ = \  -\frac{4\pi G}{e^{2}}\big(q_1^2 \ + \ q_2^2\big)\, , \nonumber \\[2mm]
&&a_{2}\  + \ \Lambda \ = \  \frac{4\pi G}{e^{2}}\big(q_1^2 \ + \ q_2^2\big)\, ,
\end{eqnarray}
which must be satisfied together with Eq.~(\ref{resenikompletAP}). In the vacuum case, these relations reduce to $ a_{2} \ = \  f_{2} = -\Lambda\,$, corresponding to the STs discussed in Sec.~\ref{const_curv}.

\vspace{3mm}

 \textbf{2)} The different but related question to the previous one concerns the GR solutions of the same problem --- namely, metric~(\ref{2+2m}) together with EM (or YM) field~(\ref{polesep}). To address this, consider augmenting the Einstein--Hilbert action
\begin{eqnarray}
&&\mbox{\hspace{-5mm}}S_{\rm EH} \nonumber \\[2mm] 
&&\mbox{\hspace{-5mm}}=  \frac{1}{16\pi G}\!\int d^{4}x\,
\sqrt{|g_{(\alpha)}||g_{(\beta)}|}
\left(R_{(\alpha)} +  R_{(\beta)} -  2\Lambda\right)\, .~~~~~
\end{eqnarray}
The EM part of Eq.~(\ref{withtrace}) is then reproduced up to an overall multiplicative constant, while the gravitational variation of the $(\alpha)$ sector yields
\begin{equation}
\left(-\tfrac{1}{2}R_{(\beta)} \ + \ \Lambda\right) g_{ij}\, .
\end{equation}
Tracing the equations in both the $(\alpha)$ and $(\beta)$ sectors and adding them together gives
\begin{equation}
-\frac{1}{2}R_{(\alpha)} \ - \  \frac{1}{2}R_{(\beta)} \ + \  2\Lambda \ = \  0\, ,
\end{equation}
which directly implies that both Ricci scalars are constant. Hence, each 2D submanifold is maximally symmetric, and the canonical reduction~(\ref{canonical}) applies: $A(r)$ and $F(x)$ must be second-degree polynomials. Substituting these forms back into (\ref{IV.87.kl}) equations forces the nonvanishing components of the EM field to be constant as well.

Consequently, the remaining field equations reduce to the same algebraic constraints~(\ref{Ealg})  obtained in the preceding discussion. The vacuum sector therefore coincides with the previous case --- as expected, since all vacuum Einstein STs are WCG solutions. For non-vacuum configurations, however, the solution space is slightly larger because the additional restriction~(\ref{resenikompletAP}) no longer applies.

\vspace{3mm}

\textbf{3)} 
Another question concerns the existence of GR solutions that are Weyl related to the full solution~(\ref{III.58.kl})-(\ref{resenikompletAP}). This is a rather involved problem, so we begin with the vacuum case in which one of the 2D sectors has constant curvature [i.e., we employ the notation of~(\ref{RDE}) or~(\ref{RDL})] and assume that $\Omega=\Omega(\vartheta)$ in the equations 
\begin{equation}
\label{Etr}
 \tilde{G}^{\mu\nu}+\Lambda \tilde{g}^{\mu\nu}=0   
\end{equation}
for the Weyl transformed metric. In this setting, it is possible to extract the differential equation
\begin{equation}
-2(\Omega')^{2} \ + \ \Omega''\,\Omega \ = \  0\, ,
\end{equation}
which in standard diagonal coordinates, cf.~(\ref{canonical}), follows from subtracting the $\chi\chi$-component from the $\vartheta\vartheta$-component. The solution is
\begin{equation}
\label{OmTh}
\Omega \ = \  (C_{1}\vartheta \ + \  C_{2})^{-1}\, .
\end{equation}
Once this form of $\Omega$ is inserted, the remaining component equations in~(\ref{Etr}) reduce to algebraic relations among the constants $C_{1}$, $C_{2}$, and $\Lambda$. For $b_{3}\neq 0$ the system admits solutions with 
\begin{eqnarray}
&&C_1 \ \text{arbitrary}, \quad C_{2} \ = \ \frac{C_{1}(b_{2}-K)}{3b_{3}}\, , \nonumber \\[2mm]
&&\Lambda \ = \  \frac{C_{1}^{2}\left(2K^{3}-3K^{2}b_{2}-27b_{0}b_{3}^{2}+b_{2}^{3}\right)}{9b_{3}^{2}}\, .
\end{eqnarray}
This Weyl transformation generically introduces singular points: it enhances the singular behavior as $\vartheta\to\pm\infty$ and produces an additional divergence at $C_{1}\vartheta + C_{2}=0$.

The solution obtained from~(\ref{RDE}) should (by the BT)  correspond to the Schwarzschild--(anti-)de Sitter geometry (or its $K$-generalizations~\cite{kniha,TBH}). Indeed, the coordinate transformation $\tilde r^{2}:=\Omega^{2}(r)$ brings the metric into a standard form by eliminating the linear term. An analogous construction applies in the case of~(\ref{RDL}). For $b_{3}=0$, equation~(\ref{tridy}) allows two possibilities, $b_2 = \pm K$. 
The second case has already been addressed in the previous discussion, as it corresponds to an Einstein space even without a Weyl transformation. For the first possibility, we obtain
\begin{eqnarray}
\label{Om1cc}
&&C_1 \ \text{arbitrary}, \quad C_{2} \ = \ \frac{C_{1}b_1}{2K}\, , \nonumber \\[2mm]
&&\Lambda \ = \  \frac{3C_{1}^{2}\left(b_1^{2} \ -  \ 4b_0K\right)}{4K}\, . 
\end{eqnarray}
No Weyl related Einstein solution exists for the general charged system (or the YM–coupled version). This follows directly from the Einstein equations together with the Weyl rescaling $\tilde{T}^{\mu \nu} = \Omega^{-6} T^{\mu \nu}$: although the functional form of $\Omega$  given by~(\ref{OmTh}) remains unchanged, substituting it into the remaining field equations (which become polynomials in $\vartheta$) shows that the highest-order terms cannot vanish unless both charges $q$'s are zero (or $\Omega$ is constant). A closely related situation was discussed in Ref.~\cite{EF}.

For the general case~(\ref{III.58.kl})–(\ref{resenikompletAP}), the system is considerably more involved than in the special case where one of the 2D sectors has a constant curvature. Nevertheless, by assuming a conformal factor of the form $\Omega=\Omega(r,x)$, we find --- see Appendix~\ref{apB} for the detailed derivation --- that
\begin{equation}
\Omega(r,x) \ = \  (a x \ + \  b r \ + \  c)^{-1} \,.
\label{98.kl}
\end{equation}
With this choice, the (\ref{Etr}) equations reduce to algebraic relations for the constants $a$, $b$, $c$ and $\Lambda$ and yield for $a_3, \: f_3\neq 0$ [with $a_1$ expressed using~(\ref{resenikompletAP})]
\begin{equation}
a \ \text{arbitrary}, \quad b \ = \  \frac{a f_3}{a_3}, \quad c \ = \  \frac{a(f_2 + a_2)}{3 a_3} \,.
\label{99.jk} 
\end{equation}
The rather lengthy expression for $\Lambda$ specified by Eq.~(\ref{Lambdafull}) can be obtained in a straightforward manner. For the cases where $a_3 = 0$ or $f_3 = 0$, the results discussed above directly apply. 
Furthermore, for conformally flat cases, these results can be further generalized (see Appendix~\ref{apB}).

With (\ref{Om1cc}), the charged solution is again prohibited due to inconsistent polynomial orders. The obtained Einstein metric corresponds to a $C$-metric (up to the linear redefinitions of $r$ and $x$ to bring it into standard form~\cite{kniha}). As before, the conformal factor $\Omega$ can vanish or diverge, potentially inducing global-structure modifications similar to, or more intricate than, the constant-curvature-part case.

\subsection{Relation to other known WCG solutions}
\label{WCGrel}

\textbf{1)}
As noted earlier, the solutions~(\ref{III.58.kl})-(\ref{resenikompletAP}) represent the charged generalization of the solutions from~\cite{HJS1} (and \cite{HJS2D}), although their authors employ a slightly different coordinate system [see their Eq.~(33)]. By introducing linear redefinitions
\begin{eqnarray}
\tilde r \ = \ \alpha_{1} r \ + \ \beta_{1},\qquad 
\tilde x \ = \ \alpha_{2} x \ + \ \beta_{2}\, ,
\end{eqnarray}
together with a constant conformal factor $\Omega$, one can eliminate one polynomial order (apart from the cubic term) and fix several coefficients by an appropriate choice of $\alpha_i$ and $\beta_i$. In our coordinate system, and including electric and magnetic charges, locally Weyl equivalent geometries appear in Ref.~\cite{C_WCG} (see also references therein), where they are identified as charged $C$-metric solutions. In that analysis, the solutions are essentially obtained by adopting an ansatz in which (\ref{III.58.kl}) with~(\ref{resenikomplet}) is multiplied by a prescribed conformal factor. (In contrast, in the present work we demonstrate that Eq.~(\ref{resenikomplet}) constitutes the \emph{most general} $2+2$ direct-product ST structure in WCG.)

Regarding the conformal factor, the charged $C$-metric solutions of Ref.~\cite{C_WCG} can be generated from~(\ref{resenikomplet}) by multiplying the metric by
\begin{eqnarray}
\Omega^{2} \ = \ (x-r)^{2}\, ,
\end{eqnarray}
which introduces an additional local  $O$-type region at $x=r$ and yields divergences in the limit $x-r\to\pm\infty$. (In cases where such divergences are already present in Eq.~(\ref{I}), the conformal transformation simply enhances them.)

Another treatment of the $C$-metric within WCG appears in Ref.~\cite{NWmass2}, which focuses on SU($N$) YM fields. Their ``$C$-metric ansatz'' again yields a solution that is (Weyl) equivalent to ours when restricted to the vacuum sector. In the non-vacuum YM case, performing a Weyl transformation to the $2+2$ direct-product ST form shows that the YM fields obtained in Ref.~\cite{NWmass2} indeed satisfy the condition derived in Sec.~\ref{YM}, namely that the sums of the square roots of $F^{q\,01}$ and $F^{q\,23}$ must remain constant. (In particular, in the notation used here, Ref.~\cite{NWmass2} contains only nonvanishing $F^{q\,23}$ components, although their specific forms differ from the example we presented, since they are not constant.) Finally, the WCG $C$-metric with conformally coupled scalar field was analyzed in \cite{WCGCcc}. The latter work also contains formulas and a discussion of its conformal relation to the GR $C$-metric, which is in agreement with our results from Sec.~\ref{GRrel}.

\vspace{3mm}

\textbf{2)} 
Multiplying the first metric in Eq.~(\ref{RDE}) by $1/r^{2}$ and introducing the new radial coordinate $\tilde r = 1/r$, we obtain
\begin{eqnarray}
&&\mbox{\hspace{-5mm}}ds^2 =   -\tilde B(\tilde r)\, dt^{2}
 \ + \  \tilde B^{-1}(\tilde r)\, d\tilde r^{2}
 \ + \  \tilde r^{2}\frac{d\zeta\, d\bar\zeta}{\left(1 + K\zeta\bar\zeta/2\right)^{2}}\, ,
\nonumber \\[2mm]
 &&\mbox{\hspace{-5mm}}\tilde B(\tilde r)  \ = \  \frac{b_{3}}{\tilde r} \ + \  b_{2} \ + \  b_{1}\tilde r \ + \  b_{0}\tilde r^{2}\, .
\end{eqnarray}
For $K=1$ this reproduces the well-known MK solution~(\ref{MK}), whereas for $K=0,-1$ it yields its topological black-hole counterparts~\cite{TBH} (including their charged generalizations). As before, the transformation is singular: it introduces the local $O$-type region at $r=0$ (corresponding to $\tilde r\to\infty$, where the geometry becomes asymptotically conformally flat in the new coordinates) and it enhances the singular behavior as $r\to\pm\infty$ (i.e., $\tilde r\to 0$). More on related topics can be found in~\cite{OnMK, BHsinWCG}.

\vspace{3mm}

\textbf{3)} 
Wormhole geometries can be generated by multiplying the full metric~(\ref{RDE}) (with $K=1$) by a function $L^{2}(r)$ that has no zeros. Using the notation of~\cite{NP}, a general static wormhole metric can be written as
\begin{eqnarray}
ds^{2}
 = -\tilde B(r)\, dt^{2}
  +  
  \tilde B^{-1}(r)\, \tilde{\xi}(r)\, dr^{2}
 +  L^{2}(r)\, d\Sigma^{2}\, ,~~~
\end{eqnarray}
%
In our case, one identifies $\tilde B(r) = B(r)L^{2}(r)$ and $\tilde{\xi}(r) = L^{4}(r)$. Since $L(r)$ is unbounded, the transformation can induce a local Petrov $O$-type region  as $L(r)\to\infty$; in particular, such a divergence may cancel an already existing singularity at $r\to\pm\infty$, since the invariant~(\ref{ID}) behaves as
\begin{eqnarray}
\tilde I \ = \ \frac{(b_{2} \ + \  3 b_{3} r \ {\color{blue}-}  \  1)^2}{12\, L^{4}(r)}\, .
\end{eqnarray}
For wormhole configurations, the 2$+$2 gauge is especially convenient, as the Bach equations in the standard gauge are highly involved (and become even more complicated in time-dependent settings~\cite{tdcervi}). 

\vspace{3mm}

\textbf{4)} Within the symmetry class in which the Euclidean sector has constant curvature, there also exist time-dependent solutions constructed in our previous work~\cite{EF}. These may be viewed as time-dependent generalizations of the MK solution and its various $K$-dependent extensions, obtained formally by promoting each constant appearing in~(\ref{MK}) to a function of the EF coordinate $w$. It is therefore natural to expect that an analogous construction (or more precisely, an analogous gauge choice) can be carried out for the full family~(\ref{III.58.kl})-(\ref{resenikompletAP}). 

In Ref.~\cite{EF}, we identified the singular Weyl transformation that maps these time-dependent geometries to the class of metrics given by~(\ref{RDE}) and determined the associated CKVFs --- which coincide with $\xi_{(1)}$ in the present notation --- that become KVFs under this transformation. Now we would like to dig a bit deeper into this topic. In particular, we ask whether some of these time-dependent solutions, expressed in generalized EF coordinates $(w,r)$, belong to a subclass that admits no static representative, i.e. whether there exist genuinely time-dependent configurations that are not globally equivalent to any static solution.

A fully general answer to this question would require to solve Eq.~(\ref{CKVF eq. tr.}) with $\xi_{(1)}$ and vanishing RHS, or in other words, one would need to determine the most general conformal factor $\Omega$ that renders $\xi_{(1)}$ to be KVF. This is, in general, a difficult task, although it can be carried out in certain simple cases. Fortunately, an explicit solution can be avoided by the following argument. One may always perform a coordinate transformation that brings the CKVF $\xi_{(1)}$ into the form $\partial_{\tilde{t}}$. In such coordinates, the metric can be written as a $\tilde{t}$-independent metric multiplied by some function $n(\tilde{t},\tilde{r})$. In Ref.~\cite{EF}, it was shown that dividing the metric (in its original coordinates) by $r^{2}$ suffices to promote $\xi_{(1)}$ to a genuine KVF. This transformation is necessarily singular. The justification for the absence of any regular transformation is as follows. A regular rescaling capable of producing another ST in the class with $\xi_{(1)}$ as a KVF would need to cancel all singularities or zeros of $n(\tilde{t},\tilde{r})$. Yet the problematic points of $n$ depend on both $\tilde{t}$ and $\tilde{r}$, whereas any admissible regular conformal factor could depend on at most one of these coordinates. Consequently, no regular conformal transformation can map the time-dependent STs under consideration to the static ones.

\section{Discussion and conclusion}
\label{diskuse}

In this paper, we have analyzed 2$+$2 direct-product STs in WCG, along with their associated conformal families (Weyl equivalence classes), and included into our analysis also EM and YM fields that preserve the same structure.
We demonstrated that, in addition to admitting two independent second-order KTs, this class of STs generically possesses (at least) two independent, commuting, non-null KVFs. These isometries were then used to cast the metric into a convenient canonical form, allowing us to obtain the general solutions~(\ref{III.58.kl})-(\ref{resenikompletAP}). For the EM case, we derived the general and unique field strengths compatible with the geometry, while for the YM case we identified the constraint that the field strength must satisfy in order to source the 2$+$2 solutions, and we provided an explicit example that works for any choice of a (simple and compact) gauge group. The results obtained generalize those of Dzhunushaliev and Schmidt~\cite{HJS1}, who previously studied the vacuum case using a similar procedure. Conceptually, our approach is closely related to the Riegert's method employed in Ref.~\cite{Riegert} to establish the analogue of BT in WCG. Riegert's results appear as special cases of our more general framework, and their connection to our results is analyzed in detail. 
Along the way, we also addressed and clarified several issues and ambiguities present in the original Riegert's work.  Apart form the latter, we devoted a substantial part of our analysis to the problem of degenerate Weyl transformations (i.e. transformations where the conformal factor has divergencies or zero points), examining the global and causal changes they can induce and proposing systems of Weyl equivalence classes and subclasses associated with such transformations. 
In this sense, we have generalized and refined BT within WCG and investigated the implications of this extension. Specifically, the content of BT in WCG is incorporated into a more general statement regarding the existence and nature of additional symmetries for any spacetime that is Weyl-related to a 2$+$2 direct-product geometry and that admits electromagnetic or YM fields compatible with this decomposition. Moreover, the 2$+$2 gauge proves particularly useful for explicit calculations, as it substantially simplifies the analysis.

We have also analyzed the basic properties of the solutions obtained (and the whole Weyl class they belong to), including geodesics, horizons, convenient coordinate systems, and curvature invariants such as the Ricci and Weyl scalars. The solutions are of Petrov type $D$ or $O$, exhibiting both scalar curvature singularities and conformally flat regions of ST.  Their relation to GR solutions was also examined, both in the special cases where the solutions coincide with GR solutions and, more generally, for GR solutions that are Weyl related to our solutions. The latter connection exists only in the vacuum case. We found that all vacuum versions of solutions~(\ref{III.58.kl})–(\ref{resenikompletAP}) can be mapped to the GR $C$-metric via a Weyl transformation, which, however, in most cases turns out to be degenerate. 
We included explicit calculations together with further analysis of constant curvature cases (which allow for a broader class of possible Weyl transformations) into our work.
This construction extends previously known results, including the connection between the MK and Schwarzschild--(anti-)de Sitter solutions~\cite{HJS3, NWmass2}, as well as the mapping of the WCG $C$-metric to the GR $C$-metric~\cite{WCGCcc}.
It also rises the question of the most general $\Omega$-mapping to Einstein spaces.
We discussed the resulting global changes and their implications for the structure of the ST. Finally, we linked our solutions~(\ref{III.58.kl})–(\ref{resenikompletAP}) to previously known results in Weyl conformal gravity, including those of Schmidt~\cite{HJS1}, Flores~\cite{NWmass2}, MK and its topological black hole analogs~\cite{MK a dalsi, TBH}, Vaidya-type solutions \cite{EF}, wormhole solutions~\cite{NP}, and the charged WCG $C$-metric~\cite{C_WCG}, highlighting the differences and potential differences introduced by degenerate Weyl transformations between these STs. Again, questions about the ``most general $\Omega$'' arise, e.g. when discussing symmetries we can ask what is most general transformation keeping KVF being KVF or transforming CKVF into KVF.

Finally, we note that several other directions for further research emerge naturally from the present work. In addition to the physical analysis of the solutions obtained here --- and of their various Weyl related counterparts --- there remains a broad and quite unexplored class of other non-vacuum configurations. 
The conditions derived for the YM field can be further specified and employed to obtain (possibly unique) solutions for particular choices of the gauge group. This may, for example, establish a direct connection with the study of hairy black holes in WCG and relate our results to works in this area~\cite{NWmass2, 1WCGSU(2), 2WCGSU(2)}. It would also be worthwhile to examine whether the structures identified in this paper persist for other types of matter sources.
For example, one could consider null dust, as in the non-vacuum solutions of Ref.~\cite{EF}, or a conformally (i.e., non-minimally) coupled scalar field, as in~\cite{WCGCcc, symbr}. Furthermore, situations in which the matter fields do not share the ST symmetries may also exhibit a substantially richer and qualitatively different behavior. A very different but still related line of investigation would concern 3$+$1 direct-product STs, which, despite also admitting two KTs (similar to the 2$+$2 case), may differ significantly in their remaining properties from the 2$+$2 STs considered here. More broadly, similar phenomena may arise in MGTs --- or even in different ST dimensions --- where direct-product geometries are expected to display related structural features. These directions appear promising and we intend to pursue them in our future work.

\begin{acknowledgments}
%
We wish to thank Petr Hajíček and Roberto Percacci  for useful discussions and comments on the manuscript.  P.J. was supported by the Czech Science Foundation Grant (GA\v{C}R), Grant No. 25-18105S.  T.L. was supported by the Grant Agency of the Czech Technical University in Prague, grant No. SGS25/163/OHK4/3T/14.
\end{acknowledgments}

\appendix

\section{List of acronyms {\color{blue}}} 
\label{apA}
$\:$ \\
BRT: Birkoff--Riegert theorem \\
BT: Birkhoff theorem \\
(C)KT: (conformal) Killing tensor \\
(C)KVF: (conformal) Killing vector field \\
EF: Eddington--Finkelstein \\
EM: electromagnetic \\
GR: General relativity \\
MGT: modified gravity theory \\
MK: Mannheim--Kazanas \\
ST: spacetime \\
WCG: Weyl conformal gravity \\
YM: Yang--Mills \\

\section{Conformal Einstein spaces } 
\label{apB}

In this Appendix we aim to find Einstein spaces conformal to our solutions. We start with the ansatz $\Omega=\Omega(r,x)$ and STs (\ref{III.58.kl})-(\ref{resenikompletAP}) where on Lorentzian part it is convenient to adopt EF coordinates (\ref{81.gh}). The $rx$-component of the (\ref{Etr}) equations yields
\begin{equation}
2 \,\Omega_{,r}\,\Omega_{,x} \ - \  \Omega_{,rx}\, \Omega \ = \ 0 \,,
\end{equation}
which admits the solution
\begin{equation}
\label{Om2Th}
\Omega(r,x) \ = \  [a(x) \ + \  b(r)]^{-1} \,.
\end{equation}
Inserting back into equations gives for $ww$-component $ b''(r)=0$, i.e. $b(r)=br+c_1$. Now (\ref{Etr}) equations can be treated as polynomials in $r$.

The function $a(x)$ can be determined, for instance, as follows. 
From the $r$-th order terms in the $rw$ component of the field equations, we obtain
\begin{eqnarray}
\!\left[ (3f_3 x \ + \  f_2 \ + \  a_2)b \ - \ 3a_3 a(x) \  \right]  b = \ 0\, ,
\end{eqnarray}
which implies either $a(x)=a x + c_2$ or $b=0$. 
If the latter option holds, the $r$-th order terms in the $xx$-component further require $a_3=0$ [or, trivially, $a(x)=0$]. 
Even in this case, combining the $xx$ and $yy$-components yields the condition $a''(x)a(x)=0$, which again restricts $a(x)$ to be at most linear. 
Consequently, in all cases we recover the form~(\ref{98.kl}). 
Substituting this result back into Eqs.~(\ref{Etr}) reduces them to polynomial equations in $r$ and $x$.

Now we distinguish particular cases: for $a_3, f_3 \neq 0$ we get (\ref{99.jk}) with cosmological constant
\begin{widetext}
\begin{equation}
\label{Lambdafull}
 \Lambda \ = \ -\frac{a^2\Big(27a_0a_3^2 \ - \  a_2^3 \ - \  9a_2f_1f_3 \ + \  3a_2f_2^2 \ + \  27f_0f_3^2 \ - \  9f_1f_2f_3 \ + \  2f_2^3\Big)}{9a_3^2}\, .   
\end{equation} 
\end{widetext}
For cases $a_3=0, \: f_3 \neq 0$ or vice versa, results (\ref{Om1cc}) are reintroduced uniquely.

The case $a_3=f_3=0$ splits into $a_2=\pm f_2$. With plus sign, these are Einstein spaces already discussed, $\Omega$ is constant. Finally, for minus sign, more possibilities arise: we can write for (\ref{98.kl})
\begin{eqnarray}
&&a, \ b  \; \;\textrm{arbitrarry}\, , \quad c=\frac{a_1b-f_1a}{2a_2}\, , \nonumber \\[2mm] 
&& \Lambda \ = \ -3a^2f_0 \ - \  3a_0b^2 \ + \  3a_1bc \ - \  3a_2c^2\, .~~~~~
\end{eqnarray}
Since these cases are conformally flat, there should be a Weyl transformation leading to flat ST. Indeed, if we fix one of the $a, \: b$ constants such that $\Lambda=0$ we get it. 

We note, however, that although we started with $\Omega(r,x)$ because it is both suitable for calculations and logical given the STs we expected to get, it need not be only one. For example if we allow dependence on $w$, we can get for the lastly mentioned (conformally flat) case a whole family of Weyl related Einstein spaces given by
\begin{widetext}
\begin{eqnarray}
&&\Omega \ = \ \frac{1}{h_1(w)r\ + \ h_2(w)}\, , \quad h_1(w)\ = \ C_1 \ + \ C_2\sin\left(\frac{\sqrt{4a_0a_2 -  a_1^2}}{2} \, w\right ) \ + \ C_3\cos\left(\frac{\sqrt{4a_0a_2-a_1^2}}{2} \, w\right), 
\nonumber \\[2mm]
&& h_2(w) \ = \ \frac{h_1(w)a_2 \ + \ 2h_1'(w)}{2a_2}\, , \quad \Lambda\ = \ \frac{3(a_1^2 \ - \ 4a_0a_2)(C_1^2 \ - \ C_2^2 \ - \ C_3^2)}{4a_2}\, .
\end{eqnarray}
\end{widetext}
Fixing one constant $C_i$ such that $\Lambda$ becomes zero again make ST flat. Other cases are nontrivial.

\end{document}